\documentclass[fontsize=11pt,a4paper]{arxiv-article}
\bibliography{not-arxiv,LargeN-largeQ-GN}

\usepackage{Definitions}

\usepackage{hyperref}
\usepackage{slashed}
\graphicspath{{./plots/}}
\allowdisplaybreaks

\newcommand{\OfficialTitle}{
	Fermionic CFTs at large charge and large~N
}
\title{\setstretch{1.4}
	{\color{Thoughtless}\textls[-20]{\OfficialTitle}}
}

\hypersetup{pdfauthor={Nicola Dondi, Simeon Hellerman, Ioannis Kalogerakis, Rafael Moser, Domenico Orlando, Susanne Reffert},pdftitle={\OfficialTitle},%
	colorlinks=true,linkcolor=ThoughtYouWere,citecolor=ThoughtYouWere,urlcolor=ThoughtYouWere,ocgcolorlinks}

\author{%
	\begin{minipage}{.94\textwidth}
		\vspace{1cm}
		\begin{center} \dosserif%
			{\small
				\textbf{Nicola Dondi}\textsuperscript{\ding{73}},
				\textbf{Simeon Hellerman}\textsuperscript{\ding{74}},
				\textbf{Ioannis Kalogerakis}\textsuperscript{\ding{73}},
				\textbf{Rafael Moser}\textsuperscript{\ding{73}},
				\textbf{Domenico Orlando}\textsuperscript{\ding{72}\ding{73}} and
				\textbf{Susanne Reffert}\textsuperscript{\ding{73}}
			}
		\end{center}
		\authorBlock{\ding{73}}{\dosserif{}%
			Albert Einstein Center for Fundamental Physics\\
			Institute for Theoretical Physics, University of Bern,\\
			Sidlerstrasse 5, CH-3012 Bern, Switzerland}
		\authorBlock{\ding{74}}{\dosserif{}%
			Kavli Institute for the Physics and Mathematics of the Universe (WPI)\\ The University of Tokyo\\ Kashiwa, Chiba 277-8582, Japan}
			\authorBlock{\ding{72}}{\dosserif{}%
			INFN sezione di Torino.\\
			via Pietro Giuria 1, 10125 Torino, Italy}
	\end{minipage}
}

\DeclareMathOperator*{\SumInt}{%
	\mathchoice%
	{\ooalign{$\displaystyle\sum$\cr\hidewidth$\displaystyle\int$\hidewidth\cr}}
	{\ooalign{\raisebox{.14\height}{\scalebox{.7}{$\textstyle\sum$}}\cr\hidewidth$\textstyle\int$\hidewidth\cr}}
	{\ooalign{\raisebox{.2\height}{\scalebox{.6}{$\scriptstyle\sum$}}\cr$\scriptstyle\int$\cr}}
	{\ooalign{\raisebox{.2\height}{\scalebox{.6}{$\scriptstyle\sum$}}\cr$\scriptstyle\int$\cr}}
}

\date{}

\begin{document}

\numberwithin{equation}{section}

\begin{titlepage}

	\maketitle

	\thispagestyle{empty}

	\vfill\dosserif{}

	\abstract{%
		\normalfont{}\noindent{}%
		We study the large-charge sector of large-$N$ fermionic \acs{cft}s in three dimensions. Depending on the model and the nature of the fixed charge, we find two types of descriptions: in terms of a superfluid or a Fermi sphere. We explicitly compute the conformal dimensions of the lowest operator of fixed charge and in the superfluid case verify the \acs{eft} predictions for the phonon spectrum. 
	}

\end{titlepage}

\setstretch{1.1}
\tableofcontents

\newpage

\setstretch{1.1}

\section{Introduction}%
\label{sec:Introduction}

The large-charge expansion (\cite{Hellerman:2015nra}, see~\cite{Gaume:2020bmp} for a review) has been very successfully applied to the critical O(2N) vector model and other bosonic models as a way of making strongly coupled \acp{cft} analytically accessible. Working in a sector of large charge allows writing an \ac{eft} as an expansion in the large charge and computing the \ac{cft} data~\cite{Badel:2019oxl,Giombi:2020enj,Antipin:2020abu,Antipin:2020rdw,Jack:2021lja,Jack:2021ziq,Monin:2016jmo,Jafferis:2017zna,Arias-Tamargo:2019kfr,Cuomo:2020rgt,Cuomo:2020thesis,Komargodski:2021zzy,Banerjee:2017fcx,Banerjee:2019jpw,Banerjee:2021bbw,Dondi:2022wli}.

In these models, the ground state at large charge spontaneously breaks the global symmetry, giving rise to a condensate and Goldstone bosons in terms of which the \ac{eft} is expressed. In the O(2N) model in $D=3$, the operator dimension of the lowest operator of charge $Q$ has the form~\cite{Hellerman:2015nra,Monin:2016jmo}
\begin{equation}\label{eq:scalingON}
	\Delta(Q) = c_{3/2} Q^{3/2} + c_{1/2} Q^{1/2} - 0.0937 +\order{Q^{-1/2}},
\end{equation}
where the first two terms are due to the large-charge ground state, while the $Q^0$ contribution stems from the Casimir energy of the conformal Goldstone. 
While the Wilsonian coefficients $c_{i}$ are not accessible within the \ac{eft}, working in a limit of large $N$ and large charge allows their computation~\cite{Alvarez-Gaume:2019biu}.
The massless spectrum over the superfluid ground state consists of a \emph{conformal} Goldstone and $N-1$ Goldstone bosons with quadratic dispersion relation:
\begin{align}
	\omega_{\text{conf}} &= \frac{1}{\sqrt{2}}p + \dots, &  \omega_{\text{non-rel}} &= \frac{p^2}{2\mu}+ \dots,
\end{align}
each paired with a massive mode with mass of order $\mu \sim \sqrt{Q}$.

\medskip
While the bulk of the large-charge literature is concerned with bosonic systems, few forays have been made into the world of fermions~\cite{Delacretaz:2021ufg,Komargodski:2021zzy,Antipin:2022naw}.\footnote{The unitary Fermi gas, described by a non-relativistic CFT, has been studied at large charge in~\cite{Favrod:2018xov,Kravec:2018qnu,Kravec:2019djc,Orlando:2020idm,Hellerman:2020eff,Pellizzani:2021hzx,Hellerman:2021qzz}.}
In~\cite{Komargodski:2021zzy} it was found that the superfluid paradigm does not apply to the free fermion case, where sectors of fixed charge are described by Fermi surfaces.

In this article, we want to further close this gap in knowledge and systematically study a number of models with four-fermion interactions in three dimensions, such as the \ac{gn} model, the chiral \ac{gn} model or \ac{njl} model, and its $SU(2) \times SU(2)$ generalization.
Similarly to~\cite{Alvarez-Gaume:2019biu}, we work also here in the limit of large number of fermion flavors, $N\to \infty$, in order to have a controlled setting.

We find that two types of qualitative behaviors are possible.
In the \ac{gn} model there is no \ac{ssb} in sectors of large baryon number, the large-N physics is the one of an approximate Fermi surface in
the strict infinite-N limit. We have not yet determined whether the
Fermi surface persists once subleading large-N corrections are included.
The four-fermi interactions of the \ac{njl} models instead allow for \ac{ssb} to occur in specific sectors with large ``chiral" charge. 
In this case, the usual mechanism is at play, and we can verify the expectations from the large-charge \ac{eft}.

We want to stress that the lack of \ac{ssb} in the \ac{gn} model at large charge at \emph{leading order in \(N\)} can be explained in principle in two different ways:
\begin{itemize}
\item the four-Fermi interaction is repulsive in every channel and the large-\(N\) ground state is a Fermi surface. This would be the first non-free example of this behavior at large charge.%
\footnote{In our convention for labeling
\(N\), the GN model is well-defined for half-integer values of \(N\) in \(D=3\).};
\item the four-Fermi interaction has an attractive channel, runs logarithmically to strong coupling, has a \ac{bcs} condensate and a gap in the fermion sector of order \(\exp(- 1 / g_{\text{eff}})\), where \(g_{\text{eff}}\) is the effective coupling at the cutoff.  Since \(g_{\text{eff}}\) should be of order \(1/N\), this means the condensate and the gap would be exponentially small and invisible to all orders in large-\(N\) perturbation theory. %
\end{itemize}

\medskip
Regardless of the presence of \ac{ssb},  finite-density ground states for critical theories on the cylinder compute the scaling dimension of certain \ac{cft} primary operators. 
This fact is a consequence of the state-operator correspondence: a primary $\mathcal{O}_Q$ with $U(1)$-charge $Q$ and scaling dimension $\Delta(Q)$ corresponds to a state $|\mathcal{O}_Q \rangle$ on a cylinder of radius $r_0$ with total charge $Q$ and energy:
\begin{equation}
	E(Q) = \Delta(Q) / r_0 .
\end{equation}
In particular, if $\mathcal{O}_Q$ is the lightest primary of charge $Q$, its corresponding state is known to minimize the combination $\braket{ \mathcal{O}_Q | \hat{H}_{\text{cyl}} - \mu \hat{Q} | \mathcal{O}_Q}$ where $\hat{H}_{\text{cyl}}$ is the \ac{cft} cylinder Hamiltonian and $\hat{Q}$ is the charge operator. A convenient way of selecting such a state is to consider the thermal \ac{cft} on $S^1_\beta \times S^2$ and study the zero-temperature limit of its grand-canonical partition function,
\begin{equation}
 Z(\beta ,\mu) = \Tr \left[ e^{-\beta (\hat{H}_{\text{cyl}} - \mu \hat{Q})} \right] \xrightarrow[]{\beta \rightarrow \infty} \braket{ \mathcal{O}_Q | \mathcal{O}_Q} e^{- \beta (E(Q) - \mu Q)}.
 \label{eq:partition_function}
\end{equation}
In the models investigated in the present work, the partition function $Z(\beta ,\mu)$ has a path-integral representation which can be computed exactly order-by-order in the large-$N$ limit. In general, it will take the form 
\begin{equation}
    Z(\beta ,\mu ) \xrightarrow[N \rightarrow \infty]{\beta \rightarrow \infty} e^{- \beta \Omega(\mu)},
\end{equation}
where $ \Omega$ is the thermodynamic potential.
Comparing with Eq.~\eqref{eq:partition_function}\footnote{The normalization of the state $\ket{ \mathcal{O}_Q}$ is irrelevant for our purposes.}, we can compute the charge and energy of the state $\ket{\mathcal{O}_Q }$ as follows:
\begin{align}
{Q} &= - \frac{\partial {\Omega}}{\partial \mu}, & {E}(Q) &= \left. {\Omega}(\mu) + \mu {Q} \right|_{\mu = \mu({Q})} .
\label{eq:this_is_basic}
\end{align}
When we perform computations on $S^1_\beta \times T^2$ %
we will use the same symbols to indicate energy and charge density, and normalize by the volume $V$ of the torus. In what follows we will not distinguish between a torus of large volume and flat space.

\bigskip

In the fermion models analysed in this paper, 
we employ the usual large-N technology and perform a Hubbard--Stratonovich transformation, introducing a complex scalar collective \ac{dof} $\Phi$, reducing 4-fermion terms to Yukawa--type interactions.
If $\Phi$ is kept non-dynamical, the conformal phase is found in the \ac{uv} and is only treatable in the context of the large-$N$ expansion.
If $\Phi$ is made dynamical, we obtain the $D=3$ \ac{uv} completion of these models, and the same conformal phase arises in the \ac{ir}.
Independently on the realization chosen, the large-charge primary $\mathcal{O}_Q$ is part of this \ac{cft} spectrum. 

In the \ac{njl} model, at leading order in $N$, we find the scaling dimension of the operator $\mathcal{O}_Q$ to be
\begin{equation}
  \frac{\Delta_{SF}(Q)}{2N} = \frac{2}{3} \pqty*{\frac{Q}{2N \kappa_0}}^{3/2} + \frac{1}{6} \pqty*{\frac{Q}{2N \kappa_0}}^{1/2} - 0.0937 + \frac{11-6\kappa_0^2}{720\kappa_0^{2}} \pqty*{\frac{Q}{2N \kappa_0}}^{-1/2} + \dots
\end{equation}
in the regime $Q/N \gg 1$, where $\kappa_0$ is a constant defined by $\kappa_0 \tanh \kappa_0 =1$.
This matches the result of the \ac{eft} superfluid description, as it was the case for the scalar $O(2N)$ model~\cite{Alvarez-Gaume:2019biu} but with different Wilsonian coefficients. %
While we haven't computed the full scaling dimension $\Delta(Q)$ to subleading order in $1/N$, we have verified that the corresponding ground state has the expected conformal Goldstone boson excitation. This guarantees that $\Delta(Q)$ will have the universal $Q^0$ contribution stemming from its Casimir energy.
In large-$N$, we also have access to the opposite regime \(Q/N \ll 1\), where one finds%
\footnote{We will refer this limit as the ``small charge'' limit even though \(Q\) is never smaller than 1.} 
\begin{equation}
	\frac{\Delta(Q)}{2N}  = \frac{1}{2}\left(\frac{Q}{2N}\right) +\frac{2}{\pi^2}\left(\frac{Q}{2N} \right)^2+ \dots \, ,
\end{equation}
which matches the standard perturbative result for the $Q$-th power of a scalar operator of charge two and dimension one. We find equivalent results for large-charge sectors of the $SU(2) \times SU(2)$ generalization of the \ac{njl} model.

Both the \ac{gn} and \ac{njl} models are expected to have interacting fixed points in any dimension \(2 < D < 4\).
Although we will not address the \(\epsilon\) expansion in the present work
or the behavior of the models in any dimension other than \(D=3\), many details of the Lagrangians of these theories
can be understood by dimensionally reducing the \ac{gn} and \ac{njl} Lagrangians from \(D=4\).
This being the case, we will often write details of the Lagrangian and discuss the symmetry groups in quasi-four-dimensional language, using the notation \(\Gamma_5\) for the four-dimensional chirality matrix and \(C_4\) for the four-dimensional charge-conjugation matrix. We also refer to chiral and axial symmetries, even though these dimensionally reduce to ordinary global symmetries in \(D=3\).
This language is intended to make it
simpler to discuss the corresponding computation in \(D=4-\epsilon\), which we hope to address in future work.
The use of four-dimensional language in the present paper should be understood in this context, and our conventions are summarised in Appendix~\ref{sec:app_fermions}.

\bigskip
In all fermionic models supporting a superfluid ground state at large charge, there is a way to understand the appearance of the condensate in intuitive physical terms. For example, in the \ac{njl} model one can perform a \ac{pg} transformation~\cite{pauli1957conservation,gursey1958relation},
\begin{equation}
    \begin{aligned}
        &\Psi \mapsto \frac{ 1}{2} \left[ (1-\Gamma_5) \Psi - (1+ \Gamma_5) C_4 \bar\Psi^T  \right] ,
        &\bar\Psi \mapsto \frac{ 1}{2} \left[ \bar\Psi (1+\Gamma_5) - \Psi^T C_4 (1- \Gamma_5) \right] ,
    \end{aligned}
\end{equation}
to a model with a Cooper-type interaction~\cite{Kleinert:1998kj,Ebert:2016ygm}.
All computations can be repeated in the Cooper model leading to the same results. The advantage of this model is that the nature of the condensate becomes obvious even in large-$N$: it is a condensate of Cooper pairs, describing a superconductor. The attractive interaction gives rise to a Cooper instability, and we effectively have again a system of condensing bosons at large charge, explaining the similarity of our results to those of the \( O(2N) \) scalar model.

\bigskip
The plan of this article is as follows. In Section~\ref{sec:models}, we introduce all the fermionic models we want to study here. In Section~\ref{sec:symmetry}, we work in flat space, which is enough to establish the presence or absence of a condensate at large charge in the various models. In Section~\ref{sec:fluctuations}, we compute the spectrum of fluctuations by studying the one-loop propagator of the scalar fields. We discuss the expected \ac{eft} spectrum and match the results to known Goldstone bosons counting rules. In Section~\ref{sec:conformalDim}, we give the conformal dimensions of the lowest operator of charge Q for the various models, both in the limit of large and small Q. We end with conclusions and an outlook in Section~\ref{sec:conclusions}. 
Appendix~\ref{sec:app_fermions} contains our conventions and notation for gamma matrices and Dirac spinors (Appendix~\ref{sec:notation}), the reducible representation we are using for 3$D$ fermions (Appendix~\ref{sec:RedRep}) and spinors on $S^1_\beta \times S^2$ (Appendix~\ref{sec:cylinder_spinor}). In Appendix~\ref{sec:Pauli-Gursey}, we discuss the Pauli--Gürsey transformation. In Appendix~\ref{sec:loop}, we collect some Matsubara sums and loop integrals and compute the one-loop propagator for the \ac{njl} model.

\section{The Models}%
\label{sec:models}

We focus on fermionic models in three-dimensional Euclidean space which admit a second-order critical phase and are computable in the large-$N$ limit, since we want to take advantage of the controlled setting the large-N limit provides. We focus on a few explicit examples with small symmetry groups \footnote{In this classification we focus on the symmetries on top of the $SO(2N)$ symmetry}. 

Our models are all obtained by deforming the free fermion \ac{cft} by a four-fermion interaction with (irrelevant) coupling $g$. 
If we assume this model to have a fundamental \ac{uv} scale $\Lambda$ (given by some fundamental lattice discretization) then at zero temperature and zero density the irrelevant coupling has a critical value $g_c^{-1} \sim \Lambda$ where we find a scale-invariant theory separating two phases in which some symmetries are spontaneously broken. 
Our goal is to study such a critical limit at finite charge density.
If we wish to take a formal continuum limit $\Lambda \rightarrow \infty$ for this type of models, we run into the issue that these models are not renormalizable, as the coupling $g$ is irrelevant. 
It is known, however, that they are renormalizable in the $1/N$ expansion~\cite{Moshe:2003xn}: the corresponding \ac{rg} flow joins the free fermion \ac{cft} in the \ac{ir} to the conformal phase at $g_c$ in the \ac{uv}. 
There is evidence that these conformal phases survive in the finite-$N$ regime, but a proper \ac{rg} treatment requires some type of \ac{uv} completion. These completions are typically found by introducing extra scalar degrees of freedom interacting with fermionic matter via a Yukawa coupling~\cite{Zinn-Justin:1991ksq}. The advantage is that Yukawa couplings are relevant in $D<4$, so these models are \ac{uv} free and strongly interacting in the \ac{ir}, where one finds the same \ac{cft} of the four-fermi models. This \ac{cft} becomes weakly coupled in $D=4-\epsilon$, allowing for perturbative computations of its conformal data (see~\cite{Antipin:2022naw} for a computation at large charge in this regime).
Working on large-$N$ models it is sufficient to consider the minimal models with only fermion matter, so we will only briefly comment on the \ac{uv} complete version.%

The models we focus on are the \ac{gn} model and the \ac{njl} model together with its minimal \( SU(2) \times SU(2) \) generalization. 
Typically, these models are investigated either in \(D = 4 - \epsilon\) and \(D = 2 + \epsilon\)  dimensions at fixed values of \(N\), or then for \(2< D< 4\) in the limit of large \(N\). 
Of course there is no natural notion of chirality in three dimensions.
The standard approach is to dimensionally reduce the four-dimensional model and keep using four-component fermions.
A Dirac fermion in \(D = 3+1\) can be decomposed into two Majorana fermions, and in \(D = 3\), a four-component fermion sits in a reducible representation so that, starting from \(N\) Dirac fermions in \(D = 4\), we obtain \(4N\) Majorana fermions in \(D=3\).
However, there exist two inequivalent two-dimensional representations of the Clifford algebra in \(D = 3\) (which we will indentify as \(\pm \gamma_{\mu } \)), so there are \(2N + 1\) possible inequivalent choices for \(4N\) Majorana fermions.
In practice there are two interesting situations (which, confusingly enough, are both called \acl{gn} model in the literature~\cite{Erramilli:2022kgp}): either all the Majorana fermions sit in the same representation and the system has a \(O(4N)\) global symmetry, or half of the fermions sit in one representation and the other half in the other, so that the global symmetry is \(O(2N) \times O(2N) \times \setZ_2\).
In the former case there \emph{cannot be parity invariance} (which is present in four dimensions) because all the spinors have the same eigenvalue for the operator \(\gamma_0 \gamma_1 \gamma_2\).
For this reason in the following we will make the latter (parity-invariant) choice, since we are interested in the dimensional reduction of the parity-invariant four-dimensional model.
Concretely, this means that we will use four-dimensional reducible gamma matrices of the form
\begin{equation}
  \Gamma_{\mu} = \sigma_3 \otimes \gamma_\mu = \begin{pmatrix}
                          \gamma_\mu & 0 \\ 0 & - \gamma_\mu
                                         \end{pmatrix}.
\end{equation}
In this way it is possible to define a $\Gamma_5$ matrix and a notion of chirality (which is actually a flavor symmetry) in three dimensions (see Appendix~\ref{sec:RedRep} for details). 

It has been shown that results from \(4 - \epsilon\) and \(2 + \epsilon\) expansions at fixed \(N\) are completely consistent with the \(1/N\) expansion for general dimension, which includes \(D = 3\)~\cite{Fei:2016sgs}. 
Moreover, the conformal phases found are known (or strongly believed, \emph{e.g.} from lattice studies~\cite{Chandrasekharan:2013aya}) to exist also in three dimensions and finite $N$.

\paragraph{Gross--Neveu model.}

The \acl{gn} model~\cite{Gross:1974jv} in the four-dimensional reducible representation we employed is described by the Lagrangian
\begin{equation}\label{eq:L-GN}
  L = \sum_{i=1}^{N} \bar \Psi_i  \Gamma^\mu \del_\mu \Psi_i - \frac{ g}{N}  \left( \sum_{i =1}^N \bar \Psi_i \Psi_i \right)^2,
\end{equation}
which has a \(O(2N)\times O(2N) \times \setZ_2\) symmetry.
We will denote its diagonal Abelian subgroup as \(U(1)_B\):
\begin{equation}
   U(1)_B : \Psi_i \to e^{i\alpha} \Psi_i.
\end{equation}
The standard large-$N$ analysis is carried out by introducing an auxiliary Stratonovich field $\sigma$ which takes the place of the $\bar{\Psi} \Psi$ bilinear. The resulting Lagrangian is
\begin{equation}
  L =  \sum_{i=1}^{N} \bar \Psi_i \left( \Gamma^\mu \del_\mu + \sigma \right)  \Psi_i  + \frac{N}{4g} \sigma^2 .
\end{equation}
The critical limit of this theory is found by neglecting the $\sigma^2$ term, and corresponds to a second-order phase transition separating phases of broken and unbroken \( \mathbb{Z}_2 \) chiral symmetry acting as $\Psi \rightarrow -\Gamma_5 \Psi$.
At finite $N$, the appropriate \ac{uv} completion is a Gross--Neveu--Yukawa theory obtained by promoting the $\sigma$ field to a dynamical field:%
\begin{equation}
    L =  \sum_{i=1}^{N} \bar \Psi_i \left( \Gamma^\mu \del_\mu + \sigma \right)  \Psi_i  + \frac{1}{2 g_Y} \partial_\mu \sigma \partial_\mu \sigma.
\end{equation}
In the \ac{ir} limit, $g_Y$ is relevant and grows large, formally reproducing the same critical action as the \ac{gn} model. The critical point is weakly coupled in the $4-\epsilon$ expansion, allowing a finite-$N$ study of the \ac{gn} critical phase. 

\paragraph{Nambu--Jona--Lasinio--type models.}

The \ac{njl} model is a time-honored model of four-Fermi interactions with continuous chiral symmetry in four dimensions.
Its Lagrangian reads
\begin{equation}\label{eq:L-NJL}
  L =  \sum_{i=1}^{N} \bar \Psi_i  \Gamma^\mu \del_\mu \Psi_i - \frac{ g}{N}  \left[  \left( \sum_{i=1}^{N} \bar \Psi_i \Psi_i \right)^2 - \left( \sum_{i=1}^{N} \bar \Psi_i \Gamma_5 \Psi_i \right)^2  \right]
\end{equation}
The $\bar{\Psi} \Gamma_5 \Psi$ bilinear is $Sp(2N)$-invariant, thus the total symmetry group is reduced with respect to the \ac{gn} model \eqref{eq:L-GN} to $[O(2N)\times O(2N)] \cap Sp(2N) = U(N)$.
On top of the \(U(1)_B\) symmetry there is an extra $U(1)_A$ that extends the \(\setZ_2\) of the \ac{gn} model, and arises when the different quartic interactions are combined exactly as in $\eqref{eq:L-NJL}$:
\begin{equation}
  \Psi_i \to e^{i \alpha \Gamma_5} \Psi_i .
\end{equation}
The critical phase of the \ac{njl} model separates two phases in which \(U(1)_A\) is either broken or unbroken.
In the former phase, a Goldstone boson arises, differently to what happens in the \ac{gn} model where the symmetry is discrete. 

Already in the original papers~\cite{Nambu:1961tp,Nambu:1961fr}, the symmetry group was generalized to a \(U(1)_B \times SU(2)_L \times SU(2)_R \) symmetry \footnote{The total symmetry of the model is $U(N) \times SU(2)_L \times SU(2)_R$.}
by considering two-flavor fermions \(\Psi_{i,f}\), \(f = 1, 2\) with the Lagrangian
\begin{equation}
  L =  \sum_{i = 1}^{N} \sum_{f = 1}^2 \bar \Psi_{i,f}  \Gamma^\mu \del_\mu \Psi_{i,f} - \frac{ g}{N}  \left[  \left( \sum_{i = 1}^{N} \sum_{f = 1}^2 \bar \Psi_{i,f} \Psi_{i,f} \right)^2 - \sum_{a=1}^3 \left( \sum_{i = 1}^{N} \sum_{f = 1}^2 \bar \Psi_{i,f} \Gamma_5 \sigma^a_{fg} \Psi_{i,g} \right)^2  \right] ,
\end{equation}
where the \(\sigma^a\) are Pauli matrices.
This model does not have an \(U(1)_A \) symmetry. The group \(SU(2)_L \times SU(2)_R \) acts on the fermion fields as
\begin{align}
  \Psi_{i,f} &\to e^{i \frac{1 + \Gamma_{5}}{2}  \omega_a^L \sigma^a_{fg}} \Psi_{i,g} & \text{and} &&   \Psi_{i,f} &\to e^{i \frac{1 - \Gamma_{5}}{2} \omega_a^R \sigma^a_{f g}} \Psi_{i,g}  .
\end{align}
This symmetry is present thanks to the pseudo-real character of \( SU(2) \), for which there is no totally symmetric symbol $d_{abc}$. 
In the present work we will focus on the \(U(1)_B \times U(1)_A \) and \(U(1)_B \times SU(2)_L \times SU(2)_R \) \ac{njl} models, referred respectively as \(U(1)\)-\ac{njl} and \(SU(2)\)-\ac{njl}, and take the large-$N$ limit using the unbroken \( SU(N) \) group factor present in both models. 
At finite $N$, these are tractable in $d=2+\epsilon$ and around $d=4-\epsilon$ via their known \ac{uv} completions.
For the \(U(1)\)-\ac{njl} this is \footnote{
For the sake of simplicity, we will suppress all global symmetry indices from now on.}
\begin{equation}
  L =  \bar \Psi \left[ \Gamma^\mu \del_\mu + \Phi \left(\tfrac{1 + \Gamma_5}{2} \right)  + \Phi^{*} \left( \tfrac{1 - \Gamma_5}{2} \right) \right] \Psi  + \frac{1}{g_Y} \del_{\mu} \Phi^{*} \del_{\mu} \Phi,
\end{equation}
with the \( U(1)_A \) chiral symmetry realized as
\begin{align}
  \Psi &\to e^{i \alpha \Gamma_5} \Psi,  & \Phi &\to e^{-2 i \alpha} \Phi .
\end{align}
As in the \ac{gn} model, the \ac{ir} limit \( g_Y \rightarrow \infty \) formally produces the critical action used also in large-$N$ analysis. The field $\Phi$ is then identified with the Stratonovich field for the complex bilinear \( \bar{\Psi} \Psi + \bar{\Psi} \Gamma_5 \Psi \). 

A similar completion is found for the \( SU(2) \)-\ac{njl} model by introducing a set of real fields \(\sigma, \pi_{a = 1,2,3} \), so that the \ac{uv} Lagrangian is
\begin{equation}
  L =   \bar \Psi \left[  \Gamma^\mu \del_\mu  + \sigma + i  \pi_a \sigma^a  \Gamma_5 
 \right] \Psi + \frac{1}{2 g_Y} (\del_{\mu} \sigma \del_{\mu} \sigma + \del_{\mu} \pi^a \del_{\mu} \pi^a).   
\end{equation}
In this model, the symmetry \(SU(2)_L \times SU(2)_R \) acts infinitesimally as
\begin{align}
  \delta_{L,R} \Psi &= i \left(\frac{1 \pm \Gamma_{5}}{2} \right) \omega_{a} \sigma^a \Psi, &
                                              \begin{cases}
                                                \delta_{L,R} \sigma = \pm \omega_a \pi_a ,\\
                                                \delta_{L,R} \pi_a = \mp \omega_{a} \sigma +  \epsilon_{abc} \pi_b \omega_c.
                                              \end{cases}
\end{align}
Equivalently, in terms of the field \(\Phi = \sigma + \pi_a \sigma^a\),  we have the following finite transformations
\begin{align}
 \begin{cases}
   \Psi \to e^{i \frac{1 + \Gamma_5}{2} \omega_a \sigma^a} \Psi ,  \\
   \Phi \to  \Phi e^{ - i \omega_a \sigma^a}  ,
 \end{cases}
 &&
    \begin{cases}
      \Psi \to e^{i \frac{1 - \Gamma_5}{2} \omega'_a \sigma^a} \Psi , \\
      \Phi \to  e^{  i \omega'_a \sigma^a}  \Phi.
    \end{cases}
\end{align}
This model was originally discussed by Nambu and Jona--Lasinio and is also known as the iso\ac{njl} model.
\paragraph{Cooper model.}

Both the \ac{gn}-type and \ac{njl}-type models exhibit a fermion-antifermion interaction. Often, especially in the context of condensed matter physics, one wants to consider a difermion interaction in order to study superconductivity (arising via Cooper pairing). Large-$N$ Fermionic models intended for the description of superconductivity at finite $U(1)_B$-charge density generally include an interaction term of the form 
\begin{equation}
    (4f)_{Cp} = \frac{g}{N} \bar \Psi C \bar \Psi^T \, \Psi^T C \Psi ,
\end{equation}
in addition to \ac{gn}-type or \ac{njl}-type interactions~\cite{Ebert:2005gza,Ebert:2016ygm}.
In particular, we are interested in the model with just the Cooper pair interaction term,
\begin{equation}
  L =  \bar \Psi \Gamma^\mu \del_\mu \Psi + \frac{g}{N} \bar \Psi C_4 \bar \Psi^T \, \Psi^T C_4 \Psi.
 \end{equation}
In our notation, we have $C_4=\Gamma_2$, see also Appendix~\ref{sec:RedRep}.
At criticality --- differently to what happens in \ac{gn} model --- this model exhibits a non-trivial solution to the gap equation at zero temperature and finite $U(1)_B$-chemical potential giving rise to a superconducting phase. \\
It turns out that this model is dual to the \ac{njl} model in the sense of~\cite{Ebert:2016ygm}. This can be seen by applying the Pauli--Gürsey transformation~\cite{pauli1957conservation,gursey1958relation,Kleinert:1998kj},
\begin{equation}
    \begin{aligned}
        &\Psi \mapsto \frac{ 1}{2} \left[ (1-\Gamma_5) \Psi - (1+ \Gamma_5) C_4 \bar\Psi^T  \right] , &
        &\bar\Psi \mapsto \frac{ 1}{2} \left[ \bar\Psi (1+\Gamma_5) - \Psi^T C_4 (1- \Gamma_5) \right] .
    \end{aligned}
\end{equation}
For a more detailed discussion of this transformation we refer to Appendix~\ref{sec:Pauli-Gursey}.
A few remarks are in order here:
\begin{itemize}
    \item Both the Cooper model and the \ac{njl} model have a $U(1)_A\times U(1)_B$ symmetry. 
    Under a \ac{pg} transformation the $U(1)_B$-chemical potential in the Cooper model is mapped into the $U(1)_A$-chemical potential of the \ac{njl} model and vice versa.
    This means that any result obtained in the \ac{njl} model at finite $U(1)_A$ chemical potential will also apply to the Cooper model at finite $U(1)_B$ chemical potential.
    \item It is possible to compute relevant quantities at the critical point of both the Cooper model and the \ac{njl} model, which should agree up to a \ac{pg} transformation. We have explicitly checked this matching for the leading-$N$ ground state energy at finite $U(1)_B$-chemical potential in the Cooper model and $U(1)_A$-chemical potential of the \ac{njl} model.
    \item The \ac{pg} transformation corresponds to a linear involution at the level of the path-integral. Therefore, it only affects the path integral measure up to a trivial rescaling.
\end{itemize}
\section{Symmetry breaking at large $N$}
\label{sec:symmetry}

Whether or not the large-charge approach leads to simplifications depends crucially on the appearance of a condensate in the large-charge sector. We therefore first discuss symmetry breaking in the various models we consider.

\subsection{Gross--Neveu model}\label{sec:GN-det}
 
As a warm-up we consider the \ac{gn} model with $2N$ three-dimensional Dirac fermions at finite temperature and finite $U(1)_B$-chemical potential $\mu$. Using the reducible representation given in Appendix~\ref{sec:RedRep} and introducing the auxiliary field $\sigma$ for the $\bar{\Psi}\Psi$ bilinear, its action reads
\begin{equation}
S = \int_{S^1_\beta \times \mathbb{R}^2} \left[  \bar{\Psi} ( \Gamma^\mu \partial_\mu - \mu \Gamma_3 + \sigma) \Psi + \frac{N}{4 g} \sigma^2   \right].
\end{equation}
The model is equipped with a cutoff scale $\Lambda$. For generic values of the parameters $(g,\beta,\mu)$ the $\sigma$ field can acquire a \ac{vev}, spontaneously breaking the discrete parity symmetry $\Psi \rightarrow - \Gamma_5 \Psi$. Under generic assumptions, the $\sigma$-\ac{vev} configuration can be taken to be homogeneous $\langle \sigma \rangle = \sigma_0$. The large-$N$ effective action for $\sigma$ can be obtained expanding around this saddle as $\sigma = \sigma_0 + \hat\sigma/\sqrt{N}$:
\begin{equation}
S_{\text{eff}} = N \left\{  \beta V \frac{\sigma_0^2}{4g} - \Tr \log \big(D^{(\mu)}\big)^{-1}    \right\} + \frac{1}{2} \left[ \Tr (D^{(\mu)} \hat\sigma D^{(\mu)} \hat\sigma) + \frac{1}{4g}  \int_{S^1_\beta \times \mathbb{R}^2} \hat{\sigma}^2 \right] + \mathcal{O}(N^{-1}),
\label{eq:Seff_GN}
\end{equation}
where we introduced the fermionic propagator at finite chemical potential\footnote{We use the notation $X = (\tau,\vec{x})$ for points on $S^1_\beta \times \mathbb{R}^2$.}
\begin{equation}
    D^{(\mu)}(X,Y) = \braket{ X | (\Gamma^\mu \partial_\mu - \mu \Gamma_3 + \sigma_0 )^{-1} | Y}.
    \label{eq:GN_prop}
\end{equation}
\subsubsection{Leading-order action and gap equation} 

The value of the condensate $\sigma_0$ is found by minimizing the zero-order action. Using the momentum-space representation of the propagator and the fermionic Matsubara summation in~\autoref{sec:loop} one finds the expression for the grand potential
\begin{equation}
\frac{ \Omega}{N} \coloneqq  \frac{\sigma_0^2}{4 g} - 2 \int^\Lambda \frac{\dd^2 p}{(2\pi)^2} \left\{ \omega_p + \frac{1}{\beta} \log \left( 1 + e^{-\beta (\omega_p +\mu)} \right) + (\mu \leftrightarrow - \mu)  \right\},
\label{eq:GN_grandpotential}
\end{equation}
where we introduced $\omega_p^2 = p^2 + \sigma_0^2$. For consistency we need to assume $\sigma_0 , \mu \ll \Lambda$. We can then introduce the critical coupling $g_c^{-1} = \Lambda/\pi$ so that when $\sigma_0 \neq 0$ the gap equation reads
\begin{equation}
0 =  \left(\frac{1}{g} - \frac{1}{g_c}\right) - \frac{1}{\pi}  \left( \sigma_0- \frac{1}{\beta} \log (1 + e^{\beta (\sigma_0+\mu)}) - \frac{1}{\beta} \log (1 + e^{\beta (\sigma_0-\mu)})    \right).
\end{equation}
Its solution $\sigma_0$ is obtained in closed form as
\begin{equation}
	e^{\beta \sigma_0} = \frac{1}{2} \left\{  e^{\beta \pi \left( \frac{1}{g_c} - \frac{1}{g} \right)} - 2 \cosh \beta \mu + \sqrt{ \left(e^{\beta \pi \left( \frac{1}{g_c} - \frac{1}{g} \right)} - 2 \cosh \beta \mu \right)^2 -4 }   \right\}.
\end{equation}
At zero chemical potential, $\mu =0$, there is a non-trivial solution in the zero-temperature limit $\beta \rightarrow \infty$  only for $g > g_c$. It reads
\begin{equation}
   \eval*{\sigma_0}_{\mu,\beta^{-1} = 0} = \pi \left( \frac{1}{g_c} - \frac{1}{g} \right).
\end{equation}
This is the known second-order quantum phase transition of the large-$N$ \ac{gn} model at $g =g_c$, separating the parity-broken and the parity-unbroken phase. 
Outside criticality $g > g_c$, this solution survives at zero temperature and finite chemical potential as long as $\mu < \mu_c = \sigma_0\big|_{\mu,\beta^{-1} = 0}$ and for $\mu > \mu_c$ parity is restored. At the quantum critical point $g=g_c$ there is no non-vanishing solution for any $\mu$. As we will discuss in Section~\ref{sec:conformalDim}, this is consistent with the fact that in the \ac{cft} the chemical potential is sourcing a parity-even primary operator. 

The zero-temperature ground state in the critical limit is then a filled Fermi sphere of massless fermions $\sigma_0=0$. The leading-order action $S^{(0)}$ gives its grand potential
\begin{equation}
\frac{\Omega}{N} = - 2 \int_{\mu<|p|<\Lambda} \frac{\dd^2 p}{(2\pi)^2} \omega_p -2 \mu \int_{|p|<\mu} \frac{\dd^2 p}{(2\pi)^2}= - \frac{\Lambda^3}{3 \pi} - \frac{\mu^3}{6\pi}.
\end{equation}
The $U(1)_B$-charge and (renormalized) energy density of this Fermi-sphere ground state are computed using Eq.~\eqref{eq:this_is_basic} and read
\begin{align}
	\frac{Q}{N} &= \frac{\mu^2}{2\pi},  & \frac{E}{N} &= \frac{1}{3\pi} \pqty*{2 \pi \frac{Q}{N} }^{3/2} .
\label{eq:densities_GN}
\end{align}
This expression is computing the leading order in the large-$Q$ expansion for the scaling dimension $\Delta_{FS}$ of the Fermi sphere operator~\cite{Komargodski:2021zzy}, which is the lightest primary of charge $Q$. In fact, there is no difference with the free fermion \ac{cft} result at this order.

\bigskip

We would like to stress that it is not clear at this point what is the implication of the absence of \ac{ssb} at leading order in the large-\(N\) limit.  Interactions between fermions are suppressed
by the coupling \(g_{\text{eff}} = \order{1/N}\) going to zero at large \(N\), so at infinite \(N\) the ground state is described by an exactly free Fermi surface.
At finite \(N\), interaction corrections to the free-fermi operator dimensions can in principle be studied in the framework of the Fermi surface \ac{eft}~\cite{Polchinski:1992ed},
with the corrections organized in a \(1/Q\) expansion corresponding to the low-energy expansion in powers of \(1/ \mu^2 R^2\), in addition to their 
large-\(N\) suppression.

One logical possibility is that the physics remais the one of a weakly-interacting Fermi surface at the lowest energies.
However, this is not the only possibility, since the Fermi surface at finite N is never exactly free, and unlike the
case of the superfluid \ac{eft}s studied in~\cite{Hellerman:2015nra,Alvarez-Gaume:2016vff}, interactions in the Fermi surface are not automatically suppressed at low energies.
As it is well-known, the effect of the four-Fermi interaction on the excitations above the Fermi surface runs logarithmically~\cite{Polchinski:1992ed},
running to strong coupling as \(g_{\text{eff}} \propto \log(\mu / E_{\text{IR}})\) if there is any attractive four-fermi channel.  This suggests that at an infrared scale of \(R^{-1} = E_{\text{IR}} \sim 
\exp( - (\text{constant}) / g_{\text{UV}} ) \mu = \exp(- (\text{constant}) \times N ) \mu\), the Fermi surface may always develop a condensate of Cooper pairs, leading
to a gap in the fermion sector of order \(E_{\text{gap}} \sim \exp(- (\text{ constant}) \times N ) \sqrt{Q} / R\).  This scenario would be somewhat generic, and would predict the low-lying large-charge operator dimensions to be described by a purely bosonic superfluid \ac{eft}, but only for ultra-large values of the charge 
exponentially large in \(Q\) so that the enhancement of the gap by \(\sqrt{Q}\) could overcome the exponential suppression in \(N\).

As we have not done an analysis of the four-fermi interaction about the Fermi surface ground state, we do not know which of these two possibilities is 
realized; we leave that as a question to be answered in the future.

\subsection{Nambu--Jona--Lasinio model}\label{sec:NJL-det}

Next, we consider the \( U(1)_A \times U(1)_B \) \ac{njl} model at finite $U(1)_A$-chemical potential. Introducing the appropriate Stratonovich field $\Phi$, the action reads
\begin{equation}
\label{eq:U(1)-NJL-chem}
  S =  \int_{S^1_\beta \times \mathbb{R}^2} \left[ \bar \Psi \left( \Gamma^\mu \partial_\mu - \mu \Gamma_3 \Gamma_5 + \Phi P_+ + \bar{\Phi} P_- \right) \Psi + \frac{N}{4 g} \abs{\Phi}^2 \right],
\end{equation}
where $P_{\pm} = (1 \pm \Gamma_5)/2$ are the chiral projectors. This chemical potential is sourcing a finite charge density for the symmetry
\begin{align}
	\Psi &\to e^{i\alpha \Gamma_5}\Psi, & \Phi &\to e^{-2i\alpha}\Phi.
\end{align}
The leading-$N$ thermodynamic potential density for constant configurations of the $\Phi$ field is found as
\begin{align}
\frac{\Omega}{N} = \frac{\abs{\Phi_0}^2}{4 g} - \int^\Lambda \frac{\dd^2 p}{(2\pi)^2} \left\{ \Omega_+ + \Omega_- + \frac{2}{\beta} \log \left( 1 + e^{-\beta \Omega_+} \right) + \frac{2}{\beta} \log \left( 1 + e^{-\beta \Omega_-} \right) \right\},
\end{align}
where we introduced the one-particle on-shell energies
\begin{equation}
    \Omega_{\pm}^2 \coloneqq \abs{\Phi_0}^2 + (\abs{p} \pm \mu)^2.
    \label{eq:disp_NJL}
\end{equation}
The novelty with respect to the \ac{gn} model discussed above is that no Fermi sphere can arise if $\Phi_0 \neq 0$, since $\Omega_{\pm} \geq 0$. If this is the case, we can neglect the thermal logarithm terms in the zero-temperature limit, and obtain
\begin{equation}
\label{eq:NJL-FermionSpectrum}
\lim_{\beta \rightarrow \infty}	\frac{\Omega}{N} = \frac{\abs{\Phi_0}^2}{4 g} - \int^{\Lambda}\frac{\dd^2p}{(2\pi)^2} \bqty*{\sqrt{(\abs{p}+\mu)^2 + \abs{\Phi_0}^2} + \sqrt{(\abs{p}-\mu)^2 + \abs{\Phi_0}^2} }.
\end{equation}
Introducing the critical coupling $g_c^{-1} = \Lambda/\pi$, the gap equation for the $\Phi$ field at zero temperature reads
\begin{equation}
0 = \frac{\del^2 \Omega}{\del \abs{\Phi_0}^2} = \frac{1}{2}\left( \frac{1}{g} - \frac{1}{g_c} \right) + \frac{1}{2\pi^2} \left[\sqrt{ \abs{\Phi_0}^2 + \mu^2} - \mu  \arctanh*( \frac{\mu}{\sqrt{ \abs{\Phi_0}^2 + \mu^2 }})  \right].
\end{equation}
Since this equation depends only on $\abs{\Phi_0}$, we can look for a real and positive \ac{vev} $\braket{\Phi} = \Phi_0 > 0$. In the critical limit $g = g_c$ there is always a non-trivial solution to the gap equation at finite chemical potential, which reads
\begin{equation}
  \Phi_0 = \mu \sqrt{\kappa_0^2-1} ,
  \label{eq:NJL_vev_flat}
\end{equation}
where \(\kappa_0\) is the solution to the equation $\kappa_0 \tanh \kappa_0 =1$, and numerically, \(\Phi_0/\mu = 0.6627\dots\).

\begin{figure}
  \centering
    \begin{footnotesize}
    \begin{tikzpicture}
      \node at (0,0) {\includegraphics[width=.75\textwidth]{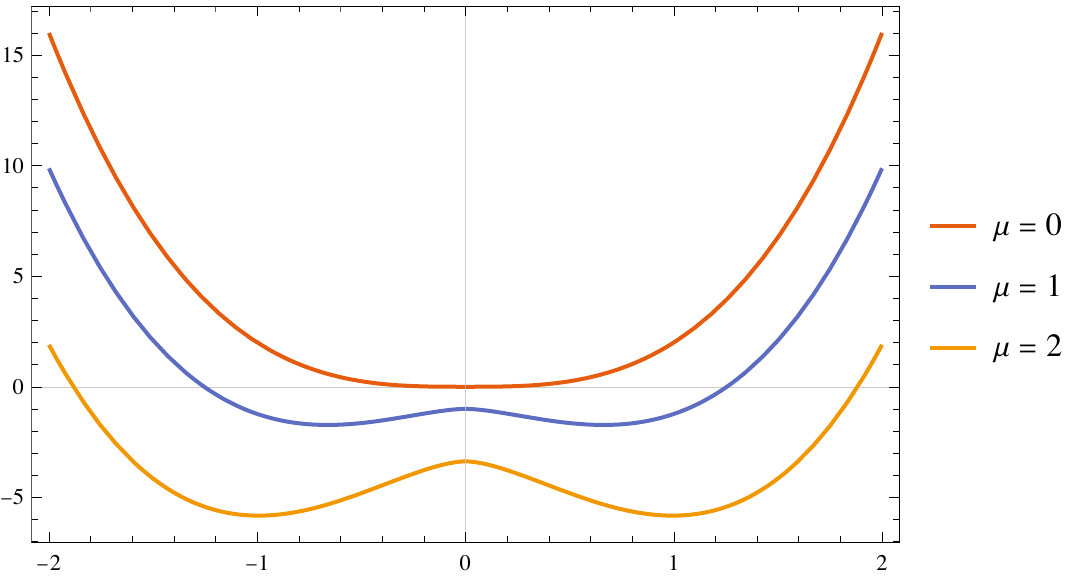}};
      \node at (-6,2.5) {\(\Omega^2\)};
      \node at (4.25,-2.75) {\(\Phi_0\)};
    \end{tikzpicture}
  \end{footnotesize}
	\caption{Leading-order in $N$ thermodynamic potential for the NJL model as function of $\Phi$ for different values of \(\mu\). For \(\mu >0 \) the minima at $\Phi\neq 0$ signal spontaneous symmetry breaking. }
 \label{fig:minima}
\end{figure}

This solution implies that the finite-$\mu$ ground state spontaneously breaks the $U(1)_A$ symmetry by giving a \ac{vev} to the auxiliary field $\Phi$, which then plays the role of the order parameter.
Symmetry restoration only occurs at $\mu=0$, as conformal symmetry prohibits the existence of a new scale separating the broken and unbroken phase. This situation becomes more explicit if one computes the renormalized potential $\Omega$ for general constant configurations of $\Phi$. Since the divergent part of $\Omega$ is $\mu$-independent, we can compute its minimal subtraction
\begin{equation}
  \begin{aligned}
	\frac{\Omega(\mu)}{N} - \frac{\Omega(0)}{N}  &= -\int\frac{\dd^2 p}{(2\pi)^2}\left[\Omega_+ + \Omega_- - 2\sqrt{p^2+ \abs{\Phi}^2}\right]\\
	&= -\frac{1}{6\pi}\left[3 \abs{\Phi}^2\mu \arctanh*(\tfrac{\mu}{\sqrt{\abs{\Phi}^2+\mu^2}}) +(\mu^2 - 2\abs{\Phi}^2)\sqrt{\abs{\Phi}^2 + \mu^2} + 2\abs{\Phi}^3\right],
  \end{aligned}
\end{equation}
and add again its renormalized value $\Omega$. This is precisely the same integral that we had solved in the computation for the \ac{gn} model and reads
\begin{equation}
   \Omega(0) = \frac{N \abs{\Phi}^3}{3\pi}.
\end{equation}
Finally the renormalized potential reads
\begin{equation}
	\frac{\Omega}{N} = -\frac{1}{6\pi}\left[3 \abs{\Phi}^2\mu \arctanh*(\tfrac{\mu}{\sqrt{\abs{\Phi}^2+\mu^2}}) + (\mu^2 - 2\abs{\Phi}^2) \sqrt{\abs{\Phi}^2 + \mu^2} \right].
\end{equation}
This potential is $U(1)_A$-invariant as expected, but has a $S^1$-worth of vacua at $\abs{\Phi} = \mu \sqrt{\kappa_0^2 -1}$. In Figure~\ref{fig:minima} we plot it for $\mathop{Im} \Phi =0$ for different values of the chemical potential.

The ground state we obtain for this model corresponds to a superfluid with $U(1)_A$-charge and energy density, respectively
\begin{align}
	\frac{Q}{N} &= \frac{\kappa_0^3 \mu^2}{2\pi} , & \frac{E}{N} &= \frac{1}{3\pi \kappa_0^{3/2}} \pqty*{2\pi \frac{Q}{N}}^{3/2}.
 \label{eq:flat_space_NJL}
\end{align}
As we will discuss in \autoref{sec:conformalDim}, this ground state energy is computing the leading order in the large-charge expansion for the scaling dimension $\Delta_{SF}$ of the operator $\mathcal{O}_{SF}$ corresponding to a superfluid ground state. No Fermi sphere arises in this sector and all the charge is contained in the superfluid. 

One might wonder whether a superfluid ground state is found also at finite $U(1)_B$-charge. In that case, instead of the eigenvalues given in Eq.~\eqref{eq:disp_NJL} one obtains
\begin{equation}
    \Omega_{\pm} = \sqrt{p^2 + \abs{\Phi_0}^2} \pm \mu.
\end{equation}
These are the same eigenvalues appearing in the \ac{gn} model, so we can conclude that in this sector no superfluid arises, but instead one has a Fermi sphere with charge and energy density as in Eq.~\eqref{eq:densities_GN}.

\subsection{$SU(2)_L \times SU(2)_R$ Nambu-Jona-Lasinio model}\label{sec:su2su2-det}

Finally, we consider the $SU(2)_L\times SU(2)_R$ \ac{njl} model. In this case we have multiple choices for charge densities to source. Let us first consider the model at finite chemical potential for the $\sigma^3$ and $\Gamma_5 \sigma^3$ generators. The critical action reads
\begin{equation}
\label{eq:SU(2)-NJL-chem}
  S =  \int_{S^1_\beta \times \mathbb{R}^2} \left[ \bar \Psi \left( \Gamma^\mu \partial_\mu + \sigma + i \pi_a \sigma^a \Gamma_5 - \left\{
        \begin{aligned} &\mu_V \Gamma_3 \sigma^3 \\
        &\mu_A \Gamma_3 \Gamma_5 \sigma^3 \end{aligned}
        \right\}
        \right) \Psi  \right].
\end{equation}
In this case, the thermodynamic potential is found to be
\begin{align}
\frac{\Omega^{V,A}}{N} = - 2 \int^\Lambda \frac{\dd^2 p}{(2\pi)^2} \left\{ \Omega_+^{V,A} + \Omega_-^{V,A} + \frac{2}{\beta} \log \left( 1 + e^{-\beta \Omega_+^{V,A}} \right) + \frac{2}{\beta} \log \left( 1 + e^{-\beta \Omega_+^{V,A}} \right) \right\},
\end{align}
where we introduced the one-particle on-shell energies 
\begin{align}
\Omega^{V}_{\pm} &= \sqrt{|\Phi_2|^2 + \left(\sqrt{(|p| + |\Phi_1|} \pm \mu_V \right)^2 }, &
\Omega^{A}_{\pm} &= \sqrt{|\Phi_1|^2 + \left(\sqrt{(|p| + |\Phi_2|} \pm \mu_A \right)^2 },
\end{align}
with $|\Phi_1|^2 = \sigma^2 + \pi_3^2,\, |\Phi_2|^2 = \pi_1^2 + \pi_2^2$. In both cases, there is no solution for the zero-temperature gap equation in which both field combinations $\Phi_1 , \Phi_2$ acquire a vev. When $\mu_V$ is turned on, one finds the solution $|\Phi_{1,0}| = 0,\, |\Phi_{2,0}| = \mu_{V,A} \sqrt{\kappa_0^2 -1}$, where $\kappa_0$ is the same transcendental number found in the $U(1)$-\ac{njl} model~\eqref{eq:NJL_vev_flat}. The same holds when $\mu_A$ is turned on, exchanging the vev for $\Phi_1, \Phi_2$. 

Since in both cases $\Omega_{\pm}^{V,A} \geq 0$ for any value of $|p|$ and $\mu_{V,A}$, no Fermi sphere arises in the zero-temperature limit, and we find the same superfluid regime of the $U(1)_B\times U(1)_A$ \ac{njl} model~\eqref{eq:NJL_vev_flat}. The only difference is an overall factor of 2 in the thermodynamic potential and ground state energy. 

We have also computed the ground states for finite left / right chemical potentials $\mu_{L,R}$, which respectively source the charge densities $ \bar{\Psi}(1 \pm \Gamma_5)\Gamma_3 \sigma^3 \Psi$. We find that the ground state is a filled Fermi sphere in both cases, just like in the \ac{gn} model and no \ac{ssb} arises.

\section{Spectrum of fluctuations}
\label{sec:fluctuations}

After having identified the large-charge ground state in the last section, we now study the spectrum of fluctuations over it, still working in flat space. In the cases with symmetry breaking, we expect, based on the superfluid \ac{eft} for the O(N) model at large charge the appearance of a \emph{conformal} Goldstone with a dispersion relation $\omega = p/\sqrt{2}  + \dots$ paired with a massive mode of order $\mu$ in all cases~\cite{Hellerman:2015nra}.
On the other hand, in the \ac{gn} model, the ground state is a filled Fermi sphere and since no Goldstone bosons arise, the \ac{eft} predictions do not apply.

\subsection{GN model}
\label{sec:gn-fluctuations}

The fluctuations around the Fermi sphere ground state can be both fermionic or bosonic. The fermionic fluctuations are of the particle-hole type, as it also happens in the free fermion critical theory. Bosonic fluctuations are instead due to the composite field $\sigma$. 
To understand their effect on the ground state, we need to compute the effective propagator of the $\sigma$ field in the unbroken phase $\sigma_0=0$ at criticality. This can be read off the \textsc{nlo} effective action $S^{(2)}_{\text{eff}}$ in Eq.~\eqref{eq:Seff_GN}. 
It is convenient to compute this action in momentum space,\footnote{We use the notation $P = (\omega_n , \vec{p})$ for momenta on $S^1_\beta \times \mathbb{R}^2$, where $\omega_n$ is the fermionic Matsubara frequency.} where 
\begin{align}
	\Tr ( D^{(\mu)} \sigma D^{(\mu)} \sigma) &= - \SumInt \dd^2 p\, \sigma(-P) \sigma(P)  \SumInt \frac{\dd^2 k}{\beta (2\pi)^2}  \Tr[D^{(\mu)}(K) D^{(-\mu)}(P-K)].
 \label{eq:doubletrace}
\end{align}
In this expression we made use of the property $D^{(\mu)}(X,Y) = -D^{(-\mu)}(Y,X) $ and the expression for the momentum-space fermionic propagator at finite density,
\begin{align}
	D^{(\mu)}(P)&= \frac{i \Gamma_\mu \tilde{P}^\mu}{\tilde{P}^2}, & \tilde{P}& = (\omega_n - i \mu, \vec{p}).
\end{align}
The one-loop integral appearing in Eq.~\eqref{eq:doubletrace} can be reduced, after some algebra, as
\begin{equation}
\vcenter{\hbox{\includegraphics[scale=1]{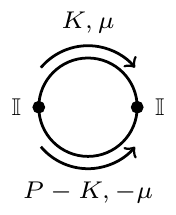}}} \, = 2 P^2 I_2 -4 I_1,
\end{equation}
where $I_1,I_2$ are two master scalar integrals computed in detail in~\autoref{sec:loop}. This result is sufficient to obtain the quadratic fluctuation action $S^{(2)}_{\text{eff}}$ in a mass-independent regularization scheme in which the critical coupling is $g_c^{-1}=0$. At zero temperature, one finds
 \begin{equation}
 	S^{(2)}_{\text{eff}} = \frac{1}{2} Tr(\sigma D^{(\mu)} \sigma D^{(\mu)}) = \frac{1}{2} \SumInt \dd^2 p \, \sigma(-P) \sigma(P) \left[ \frac{\sqrt{P^2}}{4} + \frac{\mu}{\pi} \right] .
 \end{equation}
The non-local action above does not describe stable bosonic fluctuations on top of the Fermi-sphere ground state. In fact, the momentum-independent part cannot be interpreted as a mass term, but rather as a decay constant of order $\sim \mu$, as it is the case in the unbroken phase of the model discussed in~\cite{Hands:1992be}.

\subsection{NJL model}
\label{sec:njl-fluctuations}

In the previous section, we have seen that fixing the axial charge in the \ac{njl} model leads to a \ac{vev} for the field $\Phi$. From the determinant in Eq.~\eqref{eq:NJL-FermionSpectrum} we see that all the fermions acquire a mass,
\begin{equation}
	m_F^2 = \mu^2 + \Phi_0^2 = \kappa_0^2\mu^2.
\end{equation}
This shows that the flavor symmetry remains unbroken, while the \( U(1)_A\) symmetry is broken. 

In the present model, the \ac{eft} prediction can be verified explicitly by computing the $1/N$ term in the expansion of the functional determinant, which gives the propagator for the fluctuations of the collective field $\hat \Phi = \hat\sigma + i \hat\pi$ over the vacuum $\Phi_0$.

To identify the spectrum of fluctuations over the condensate, we compute the inverse propagator of the field $\hat\Phi$ with one fermion loop. The Lagrangian around the vacuum $\braket \Phi = \Phi_0 $ is
\begin{equation}
	L_{\Phi_0} =  \bar\Psi \Gamma_\mu \partial^\mu \Psi + \Phi_0\bar\Psi\Psi - \mu \bar\Psi\Gamma_3\Gamma_5\Psi + \frac{1}{ \sqrt{N}} ( \hat\sigma \bar\Psi\Psi+ i\hat\pi \bar\Psi\Gamma_5\Psi ) .
\end{equation}
The fermion propagator is
\begin{equation}
\begin{split}
	D^{(\mu,\Phi_0)}( P) & = (-i\slashed{P} +\Phi_0 -\mu\Gamma_3\Gamma_5)^{-1}\\
	& = \frac{
    \left( \omega^2 + k^2 + \Phi_0^2 - \mu^2 + 2 \mu (i \omega \Gamma_3 + \Phi_0 ) \Gamma_3 \Gamma_5 \right) 
    }{\left(\omega^2 + \Phi_0^2 + (\mu + k)^2 \right)\left(\omega^2 + \Phi_0^2 + (\mu - k)^2 \right)} \left( i\slashed{P} +\Phi_0 -\mu\Gamma_3 \Gamma_5\right) ,
\end{split}
\end{equation}
where ${P} = (\omega, \vec{p})$ and $\slashed{P} = \Gamma^\mu P_\mu$.
Thanks to the absence of a Fermi sphere ground state, we can work directly in the $\beta \rightarrow \infty$ limit and drop all temperature-dependent contributions. This also amounts to considering Matsubara frequencies as continuous $\omega_n \rightarrow \omega$. 
Note that the propagator is properly antisymmetric $D^{(\mu,\Phi_0)}( -P) = -D^{(-\mu,-\Phi_0)}( P)$.

Using this fermionic propagator we can obtain the inverse propagator for the scalar fluctuations in terms of the following momentum space integrals: 
\begin{align}
\vcenter{\hbox{\includegraphics[scale=1]{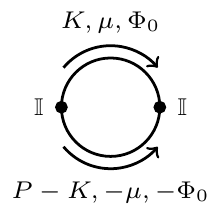}}} \, &=	D^{-1}_{\sigma\sigma}(P) = -\int \frac{\dd^3k}{(2\pi)^3} \Tr\left[D^{(\mu,\Phi_0)}(K)D^{(-\mu,-\Phi_0)}(P-K) \right],\\
\vcenter{\hbox{\includegraphics[scale=1]{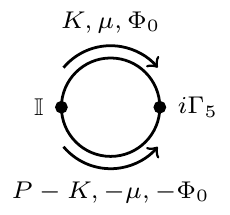}}} \, &=	D^{-1}_{\sigma\pi}(P) = -i\int \frac{\dd^3k}{(2\pi)^3}\Tr\left[D^{(\mu,\Phi_0)}(K)\Gamma_5 D^{(-\mu,-\Phi_0)}(P-K) \right],\\
\vcenter{\hbox{\includegraphics[scale=1]{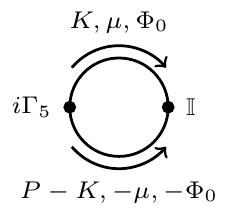}}} \, &=	D^{-1}_{\pi\sigma}(P) = -i\int \frac{\dd^3k}{(2\pi)^3}\Tr\left[ \Gamma_5 D^{(\mu,\Phi_0)}(K) D^{(-\mu,-\Phi_0)}(P-K) \right] ,\\
\vcenter{\hbox{\includegraphics[scale=1]{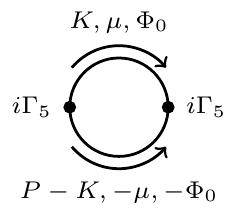}}} \, &=	D^{-1}_{\pi\pi}(P) = \int \frac{\dd^3k}{(2\pi)^3}\Tr\left[\Gamma_5 D^{(\mu,\Phi_0)}(K)\Gamma_5 D^{(-\mu,-\Phi_0)}(P-K) \right].
\end{align}
All these integrals can be conveniently expanded in the regime of interest $(P/\mu)\ll1$.
The zeroth order corresponds to the $P=0$ result and needs to be regularized by subtracting the $\mu=0$ result, since the divergence is $\mu$-independent (see Appendix~\ref{sec:njl-computations}). 
The result is
\begin{equation}
	D^{-1}(P)\Big|_{\mathcal{O}(0)} =  \begin{pmatrix}
		D^{-1}_{\sigma\sigma}(0) & D^{-1}_{\sigma\pi}(0)\\
		D^{-1}_{\pi\sigma}(0) & D^{-1}_{\pi\pi}(0) 
	\end{pmatrix} = \frac{\kappa_0\mu}{\pi}
    \begin{pmatrix}
		1 & 0\\
		0 & 0 
	\end{pmatrix},
\end{equation}
where $\kappa_0$ again satisfies $\kappa_0 \tanh \kappa_0 =1$, with $\Phi_0 = \mu \sqrt{ \kappa_0^2 -1}$. This is consistent with a massive and a massless mode. Beyond the zeroth order no regularization is necessary. 
To linear order in P we find
\begin{equation}
	D^{-1}(P) \Big|_{\mathcal{O}(P/\mu)} =  \frac{\kappa_0 \omega}{2\pi } \begin{pmatrix}
	    0 & -1 \\  1 & 0
	\end{pmatrix}.
\end{equation}
Finally, the quadratic order is computed to be
\begin{equation}
	D^{-1}(P) \Big|_{\mathcal{O}(P^2/\mu^2)} = \begin{pmatrix}
        \frac{(2 \kappa_0^2 -1) \omega^2 }{ 12 \pi \kappa_0 (\kappa_0^2 -1) \mu} + \frac{ (3\kappa_0^6 - 2\kappa_0^4 - 2 \kappa_0^2 + 2) p^2 }{ 24 \pi \kappa_0^3 (\kappa^2 -1) \mu}  & 0  \\
        0   &  \frac{ \kappa_0 \omega^2 }{ 4 \pi(\kappa_0^2 -1) \mu} + \frac{ \kappa_0^3 p^2 }{ 8\pi (\kappa_0^2 -1 ) \mu } 
    \end{pmatrix} .
\end{equation}
The dispersion relations of the two modes come from the zero of the inverse propagator given by the matrix

\begin{equation}
	D^{-1}(P) =  \begin{pmatrix}
        \frac{\kappa_0\mu}{\pi}+ \frac{2 \kappa_0^2 \left(2 \kappa_0^2-1\right) \omega ^2+\left(3 \kappa_0^6-2 \kappa_0^4-2 \kappa_0^2+2\right) p^2}{24 \pi  \kappa_0^3 \left(\kappa_0^2-1\right) \mu } &  - \frac{\kappa_0}{2\pi } \omega \\
     \frac{\kappa_0}{2\pi } \omega & \frac{2 \kappa_0 \omega^2 + \kappa_0^3 p^2 }{8 \pi ( \kappa_0^2 -1) \mu }
	\end{pmatrix} + \mathcal{O}(P^3/\mu^3),
\end{equation}
and read
\begin{align}\label{eq:U(1)-spectrum}
    \omega_{1}^2 &= - \frac{1}{2}p^2 +\dots ,\\
    \omega_{2}^2 &= - 12 \frac{ \left( \kappa_0^2 -1\right) \kappa_0^4 }{ \left( 2 \kappa_0^2-1 \right) } \mu^2 - \frac{\left(5 \kappa_0^6-5 \kappa_0^4-\kappa_0^2+2\right) }{ 2\kappa_0^2 (2 \kappa_0^2 -1) } p^2 +\dots
\end{align}
From this result, we recognize the expected conformal Goldstone and a radial-type mode of mass of order \(\order{\mu}\). This is one of the main results of the present works, and shows that large-charge superfluid \ac{eft}s can apply to this sector. 
We have performed the same computation for the $SU(2)$-\ac{njl} at finite chemical potentials $\mu_{A,V}$ and found the same spectrum of \eqref{eq:U(1)-spectrum} with two extra degenerate gapped modes with dispersion relation
\begin{equation}
    \omega ^2 = - 4  \kappa_0^2  \mu^2 - \frac{ \left( \kappa_0 ^2-1 \right) p^2}{ \kappa_0^2 } + \dots 
\end{equation}
Based on this result, we expect the conformal Goldstone \ac{eft} to describe the fluctuations on top of the \ac{ssb} ground state for both the $U(1)$-\ac{njl} and the $SU(2)$-\ac{njl} model. 

\subsection{Symmetry breaking patterns and Goldstones}
\label{sec:SBB-patterns}

The analysis performed in the previous section lead to the conclusion that the usual large-charge \ac{eft} of superfluid \acp{gb} describes both the ground states studied in the $U(1)$-\ac{njl} and in the $SU(2)$-\ac{njl} models.
Thus, the large-charge sectors of these two theories are identical and no \ac{gb} with quadratic dispersion relation is found.
This fact is consistent with the known counting rules for \acp{gb}. Let us consider the formal limit $R \rightarrow \infty$ of the cylinder so that we can label spacetime generators using a flat space notation.
The actions \eqref{eq:U(1)-NJL-chem} and \eqref{eq:SU(2)-NJL-chem} at zero chemical potentials and at criticality have total symmetry
\begin{equation}
  SO(4,1)_{\text{conf}} \times SU(N)  \times U(1)_B \times
  \begin{cases}
    U(1)_A & \text{($U(1)$-\ac{njl})}\\
    SU(2)_L \times SU(2)_R & \text{($SU(2)$-\ac{njl}).} 
  \end{cases}
\end{equation}
Introducing the respective axial chemical potentials $\mu_A$, reduces this symmetry to
\begin{equation}\label{eq:leftover-symm}
\mathbb{R}_\tau \times SO(3)_{\text{rot}} \times SU(N) \times U(1)_B \times
\begin{cases}
U(1)_A & \text{($U(1)$-\ac{njl})}\\
U(1)^{(3)}_{A} \times U(1)^{(3)}_B & \text{($SU(2)$-\ac{njl}),}
\end{cases}
\end{equation}
where $\mathbb{R}_\tau$ is cylinder-time translation and $SO(3)_{\text{rot}}$ are the isometries of the sphere, and $U(1)_{A,B}^{(3)}$ are the global phase symmetries generated by $\Gamma_5 \sigma^3$ and $\sigma^3$ respectively (the Cartans of $SU(2)_L \times SU(2)_R$).
In both cases the ground state arising at large-$N$ breaks spontaneously a linear combination of \(\setR_{\tau}\) and the $U(1)_A$ factor.
This is the typical setting in which spontaneous symmetry breaking of a global symmetry group occurs in a non-Lorentz invariant theory \footnote{Strictly speaking \eqref{eq:leftover-symm} is the Euclidean cylinder equivalent of a theory which is not invariant under Lorentz-boost.}. In this setting the counting of \acp{gb} is non-trivial~\cite{Nielsen:1975hm,Watanabe:2011dk,Watanabe:2012hr,Hidaka:2012ym}, see also \cite{Watanabe:2019xul} for a recent review.
The upshot is that the number of broken generators, $n_{BG}$, is related to the number of \acp{gb} with linear disperions relation, $n_A$, and the number of \acp{gb} with quadratic disperions relation, $n_B$, as follows:
\begin{equation}
n_A + 2 n_{B} = n_{BG}.
\end{equation}
Since both in the $U(1)$-\ac{njl} and $SU(2)$-\ac{njl} models we have $n_{BG} = 1$, only one \ac{gb} with linear dispersion relation can appear: the known conformal superfluid phonon.
The only difference between the two models is found in the gapped sector: the ground state fluctuation in the $SU(2)$-\ac{njl} also contains two degenerate gapped modes combined in a complex scalar charged under the unbroken $U(1)_B^{(3)}$.

\section{Conformal dimensions and local CFT spectrum}%
\label{sec:conformalDim}

So far we have considered four-fermion critical models in flat space, studying their finite-density properties in a large spatial box of volume $V$ (or analogously, on a torus of volume $V$). When instead the system is put on a cylinder $\mathbb{R} \times S^2$, the energy of the finite-density ground state corresponds to the scaling dimension of a specific primary of the \ac{cft} living at the critical point according to the relation $\Delta = r_0 E_{\text{g.s}}$, where $r_0$ is the sphere radius. 

Fermionic models can be mapped on $\mathbb{R} \times S^2$ via a Weyl transformation, see Appendix~\ref{sec:cylinder_spinor}. The upshot is that the finite-density kinetic term for fermions reads
\begin{align}
	S &= \int_{\mathbb{R} \times S^2} \left[  \bar{\Psi} ( \slashed{D} - \mu \Gamma_\tau + \sigma) \Psi  \right] & \text{with} && Q &= \int_{S^2} \bar{\Psi} \Gamma_\tau \Psi,
\end{align}
where $\slashed{D}$ is the Dirac operator on $\mathbb{R} \times S^2$.

\subsection{GN model}

In the large-$N$ \ac{gn} model the thermodynamic grand potential density on $S^1_\beta \times S^2$, analogous to the one in flat space in Eq.~\eqref{eq:GN_grandpotential}, is found to be
\begin{align}
\frac{\Omega}{N} = \frac{\sigma_0^2}{4g} - \frac{2}{(4 \pi r_0^2)}  \sum_{j = \frac{1}{2}} (2j+1) \left\{  \sqrt{\omega_j^2 + \sigma_0^2} + \text{thermal contributions}  \right\},
\end{align}
where $\omega_j = (j+1/2)/r_0$ are the eigenvalues of the Dirac operator on $S^2$. It is easy to see that the gap equation does not admit a non-trivial solution $\sigma_0 \neq 0$ at criticality $g = g_c$ for any value of $\mu$ at zero temperature, exactly as in flat space.%
\footnote{The sums need to be regularized by a smooth cutoff function such as $e^{-\omega_j/\Lambda}$ in order to preserve diff-invariance.}
The zero-temperature grand potential then reduces to
\begin{equation}
	\frac{\Omega}{N} = - \frac{1}{2\pi r_0^2} \left\{ \sum_{\omega_j > \mu} (2j+1)\omega_j + \mu \sum_{\omega_j < \mu } (2j +1) \right\}
\end{equation}
and describes a Fermi-sphere ground states for massless fermions. 
The charge and energy density of the Fermi sphere are readily found using Eq.~\eqref{eq:this_is_basic}:
\begin{align}
\frac{Q}{N} &= \frac{1}{2\pi r_0^2} \floor{\mu r_0} (   \floor{\mu r_0} +1  ), & \frac{E}{N} &= \frac{1}{6 \pi r_0^3}   \floor{\mu r_0} (  \floor{\mu r_0} +1  ) ( 2 \floor{\mu r_0} +1 ).	
\end{align}
The floor function implements the fact that on the cylinder energy levels are discretized, and for a filled Fermi sphere a chemical potential value in between two energy levels corresponds to the same filled Fermi sphere. The formal macroscopic limit $r_0 \rightarrow \infty$ reproduces the flat space results in Section~\ref{sec:GN-det} and is analogous to the large-chemical-potential limit.

This ground state corresponds to the scalar primary operator $\mathcal{O}_{FS}$ which was constructed explicitly in~\cite{Komargodski:2021zzy} in the free fermion \ac{cft}, which is parity-even. This is the lowest primary of charge $Q$ even if the \ac{cft} is an interacting one, due to the fact that the auxiliary field $\sigma$ does not condense. The same conclusion is reached for the \ac{njl} model at finite $U(1)_B$ chemical potential. The $U(1)_B$-charge and scaling dimension of $\mathcal{O}_{FS}$ are found to be, respectively,
\begin{align}
	\frac{Q}{2N} &=\floor{\mu r_0} (  \floor{\mu r_0} +1  ), &
 \frac{\Delta_{FS}}{2N} &=  \frac{1}{3}  \floor{\mu r_0} (  \floor{\mu r_0} +1  ) ( 2 \floor{\mu r_0} +1 ) =  \frac{ Q}{ 6N} \sqrt{ \frac{ 2 Q}{N} + 1} .
 \label{eq:fermi_sphere_scaling}
\end{align}
Here we have normalized $Q,\Delta_{F}$ by $2N$, the total number of three-dimensional Dirac fermions. The scaling dimension $\Delta_{FS}$ is shown in Figure~\ref{fig:CGN_scaling} as a function of $Q$. The $Q \rightarrow \infty $ asymptotics can be systematically computed, and the first few orders are
\begin{align}
	\frac{\Delta_{FS}}{N} &=
\frac{2}{3} \pqty*{\frac{Q}{2N}}^{3/2} + \frac{1}{12} \pqty*{\frac{Q}{2N}}^{1/2} - \frac{1}{192} \pqty*{\frac{Q}{2N}}^{-1/2} + \order*{\pqty*{\frac{Q}{2N}}^{-3/2}} , & \frac{Q}{2N} &\rightarrow \infty.
\end{align}
For this large-charge sector, we do not have Goldstone fluctuations describing new primaries with $\sim \mathcal{O}(1)$ gap from the Fermi sphere primary. Instead, we have particle-hole excitations creating new (generally spinful and fermionic) primaries with same charge and gap $\delta \Delta\sim \mathcal{O}(1)$.
As discussed at the end of Section~\ref{sec:gn-fluctuations}, fluctuations of the $\sigma$ field cannot consistently be used to describe new primaries with $\sim \mathcal{O}(1)$ gap from the Fermi-sphere ground state. As this is a large-$N$ result, it would be interesting to see if this effect persists for finite $N$ by and if there can be a description in terms of a Fermi liquid for the \ac{eft}.

\subsection{NJL model}

In the \( U(1)_B \times U(1)_A \) \ac{njl} model, the ground state at finite $U(1)_A$-chemical potential corresponds to a superfluid. When the \ac{cft} is placed on $S^1_\beta \times S^2$, this will compute the scaling dimension $\Delta_{SF}$ of a corresponding primary $\mathcal{O}_{SF}$. Here we treat explicitly the $U(1)_B \times U(1)_A$ \ac{njl} model, keeping in mind that the leading-\(N\) result for the \(SU(2)_L \times SU(2)_R\) model in Section~\ref{sec:su2su2-det} is obtained by replacing \(N\) by \(2N\).

\begin{figure}
\centering
\includegraphics[scale=0.6]{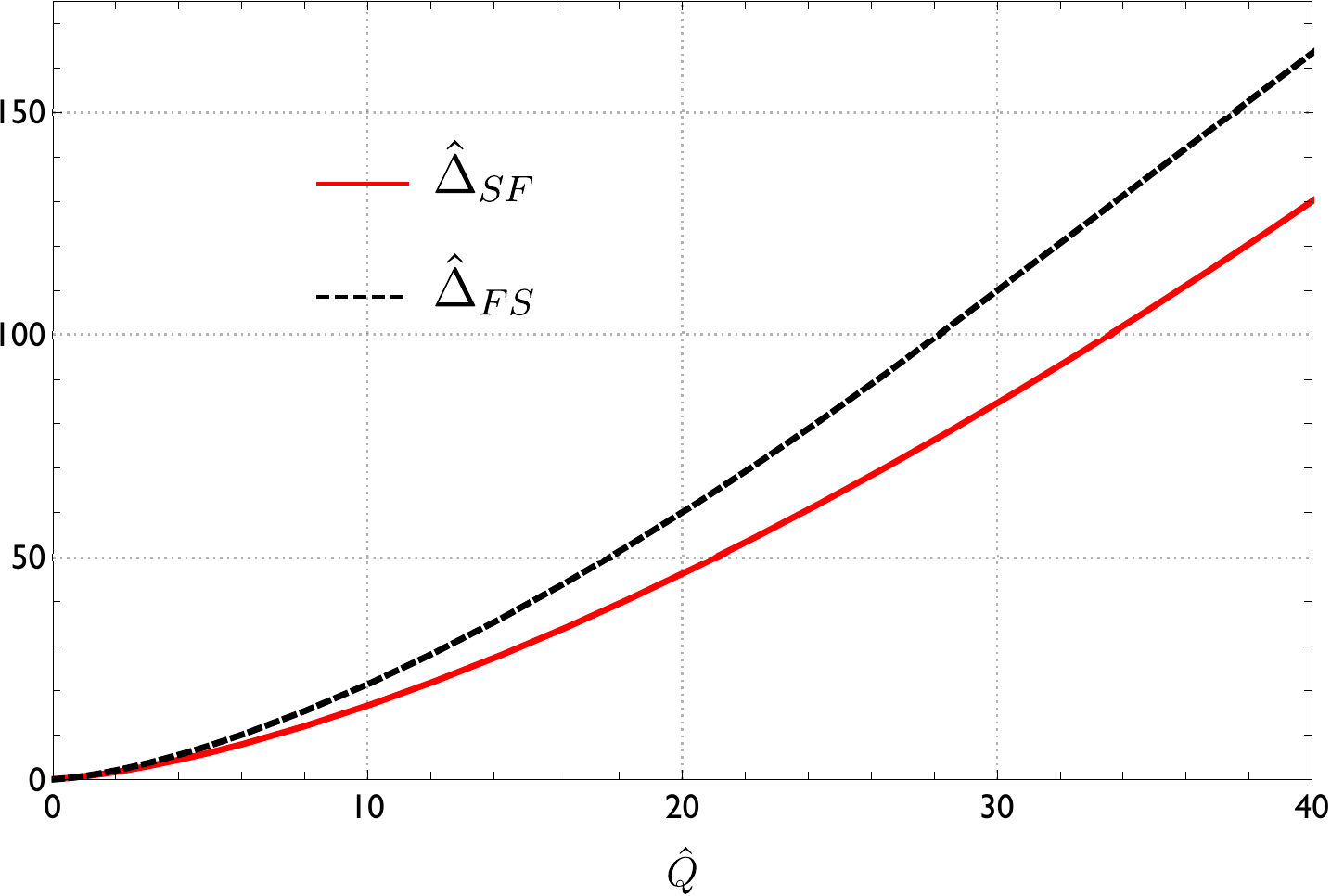}
\caption{Scaling dimension ${\Delta}_{FS}$ of the Fermi sphere primary $\mathcal{O}_{FS}$ in the GN model and NJL model at finite $U(1)_B$-charge density compared with the scaling dimension ${\Delta}_{SF}$ of the superfluid operator $\mathcal{O}_{SF}$ in the NJL model at finite $U(1)_A$-charge density.}
\label{fig:CGN_scaling}
\end{figure}

At criticality, the zero-temperature thermodynamic potential reads
\begin{align}
\frac{\Omega}{N} &= - \frac{1}{(4 \pi r_0^2)}  \sum_{j = \frac{1}{2}}^\infty (2j+1) \left\{  \Omega_+ + \Omega_-  \right\}, & \Omega_{\pm}^2 &= \abs{\Phi}^2 + (\omega_j \pm \mu)^2,
\end{align}
where $\omega_j$ are the Dirac eigenvalues on the sphere. Following the same procedure as before, we determine the charge and scaling dimension of the operator $\mathcal{O}_{SF}$ to be
\begin{align}
\frac{Q}{2N} &= \frac{1}{2}\sum_{j=\frac{1}{2}}^\infty (2j+1)\left\{ \frac{\omega_j+\mu}{\Omega_+} - \frac{\omega_j-\mu}{\Omega_-}  \right\},  \\
\frac{\Delta}{2N} &= - \frac{r_0}{2} \sum_{j=\frac{1}{2}}^\infty (2j+1) \left\{ \Omega_+ + \Omega_- \right\}  + (\mu r_0) \frac{Q}{2N}.
\label{eq:NJL_scalingdim}
\end{align}
Moreover, the auxiliary field $\Phi$ needs to be evaluated on the solution $\Phi_0$ of the gap equation,
\begin{equation}
G \coloneqq \frac{1}{2 r_0} \left. \sum_{j=\frac{1}{2}}^\infty (2j+1) \left\{ \frac{1}{\Omega_+} + \frac{1}{\Omega_-}  \right\} \right|_{\Phi= \Phi_0} = 0.
\label{eq:gap_NJL}
\end{equation}
While the charge is finite, the scaling dimension and the gap equation, on the other hand, need to be regularized. This can be done, for example, by removing the leading divergence in the sums and adding them back in a $\zeta$-function regulated form. This leads to the following regulated expressions:
\begin{align}
G^{\text{reg}} &= \frac{1}{2} \sum_{j = \frac{1}{2}}^\infty  \left\{ (2j+1) \frac{1}{r_0}\left[\frac{1}{\Omega_+} + \frac{1}{\Omega_-} \right] - 4   \right\}  + 2 \zeta(0), \\
\frac{\Delta^{\text{reg}}}{2N} &= - \frac{1}{2} \sum_{j=\frac{1}{2}}^\infty \left\{ (2j+1) r_0 \left[ \Omega_+ + \Omega_- \right] - 4 r_0^2\omega_j^2 -2 r_0^2\Phi^2   \right\} - r_0^2 \Phi^2 \zeta(0) + (\mu r_0) \frac{Q}{2N}.
\end{align}
These regulated expressions can be computed numerically for different values of the charge $Q$, as the infinite sum now converges. The result is plotted in Figure~\ref{fig:CGN_scaling} together with the scaling dimension $\hat{\Delta}_{FS} = \Delta_{FS}/2N$ for the operator $\mathcal{O}_{FS}$ appearing both in the \ac{gn} and \ac{njl} model at finite $U(1)_B$ chemical potential. 

The first few terms in the asymptotic expansion of $\Delta_{SF}$ in the regimes $\frac{Q}{N} \gg 1$ and the $\frac{Q}{N} \ll 1$ can also be computed analytically. This is done in the following sections.

\subsubsection{Large-Q regime}%
\label{sec:largeQ}

At finite $U(1)_A$ charge density, the problem has three dimensionful quantities, $\mu$, $\sigma_0$ and $r_0$, but in the above equations only the two dimensionless ratios $r_0 \mu,\, r_0 \Phi_0$ appear. Here, we are interested in the large-charge limit in which $\mu$ is the dominant scale, so that $r_0 \mu \gg 1$.
In this limit, the solution of the gap equation has the expansion
\begin{equation}
	\Phi_0 r_0 = \sqrt{\kappa_0^2-1}\left(\mu r_0 + \frac{\kappa_1}{\mu r_0}+ \frac{\kappa_2}{(\mu r_0)^3}+\dots\right).
 \label{eq:ansatz}
\end{equation}
Note that $r_0$ enters via the curvature of the sphere which has dimension $[\mathop{Ric}]=2$ and this is why we have an expansion in $r_0^2$.
In order to determine the coefficients $\kappa_i$, we need to solve the gap equation:
\begin{equation}
G =	G_r + G_d = 0, 
\end{equation}
where we have separated the regular and the divergent parts:
\begin{align}
	G_r &=  \sum_{\ell=1} \ell \left[\frac{1}{\sqrt{(\ell + \mu r_0)^2+ (\Phi_0 r_0)^2}} + \frac{1}{\sqrt{(\ell - \mu r_0)^2+ (\Phi_0 r_0)^2}} - 2 \frac{1}{\sqrt{\ell ^2+ (\Phi_0 r_0)^2}} \right] ,\\
	G_d &=  2 \sum_{\ell=1} \ell  \frac{1}{\sqrt{\ell ^2+ (\Phi_0 r_0)^2}},
\end{align}
where we also shifted the summation in Eq.~\eqref{eq:gap_NJL} as $j = \ell - 1/2$.
The regular part can be written as an asymptotic expansion in $\mu r_0 \gg 1$ using the Euler--Maclaurin formula:
\begin{equation}
	G_r = 2 \mu r_0 \left[ - \kappa_0 + \sqrt{\kappa_0^2-1} + \arccoth(\kappa_0)\right] + \frac{1}{6 \mu r_0}\left[-\frac{1}{\kappa_0}-12\kappa_0\kappa_1 +\frac{1+12(\kappa_0^2-1)\kappa_1}{\sqrt{\kappa_0^2-1}}  \right]+\dots
\end{equation}
For the divergent part, we perform $\zeta$-function regularization\footnote{These expansion coincide with the cutoff-independent part in smooth cutoff regularization.}:
\begin{align}
	G_d &= 2 \eval*{ \sum_{\ell=1}^\infty \ell (\ell^2+r_0^2\Phi_0^2)^{-s}}_{s=1/2}= \eval*{ \frac{2}{\Gamma(s)}\int_0^\infty \frac{\dd{t}}{t}t^s\sum_{\ell=1}^\infty \ell e^{-(\ell^2+r_0^2\Phi_0^2) t} }_{s=1/2}.
\end{align}
In the limit $\mu r_0 \gg 1$, we also have $\Phi_0 r_0 \gg 1$ as is evident from the flat-space result in Eq.~\eqref{eq:NJL_vev_flat}. In the $\zeta$-function regulated expression we can then obtain an expansion for $\Phi_0 r_0 \gg 1$ by expanding the integrand around $t=0$, using the standard formula 
\begin{equation}
\sum_{\ell=1}^\infty \ell e^{-\ell^2t} = \frac{e^{-t}}{12}\left(2t + 5 +\frac{6}{t} + \dots \right),
\end{equation}
resulting in
\begin{equation}
	G_d = -2 \Phi_0 r_0 - \frac{1}{6\Phi_0 r_0} - \frac{1}{120 (\Phi_0 r_0)^3} + \dots.
\end{equation}
If we now insert the ansatz \eqref{eq:ansatz} we can solve the equation $G_d + G_r = 0$ order by order in $r_0 \mu$ and solve for the coefficients $\kappa_i$. 
We then find that the first coefficient, $\kappa_0$, satisfies the same irrational equation we had found in Section~\ref{sec:symmetry}, and appears in all coefficients $\kappa_{i >0}$:
\begin{align}
	\kappa_0 \tanh \kappa_0 &= 1, &
	\kappa_1 &= -\frac{1}{12 \kappa_0^2}, &
	\kappa_2 &= \frac{33-16 \kappa_0^2}{1440 \kappa_0^6}, & &\dots
 \label{eq:coeffs}
\end{align}
We apply the same strategy to compute the divergent sum appearing in the scaling dimension~\eqref{eq:NJL_scalingdim}, which coincides with the thermodynamic potential $\Omega/(2N)$: we divide it into the two contributions
\begin{align}
	\Omega_r &= -2N\sum_{\ell = 1} \ell 
	\left[ \sqrt{(\ell +\mu r_0)^2 + \Phi_0^2r_0^2} + \sqrt{(\ell -\mu r_0)^2 +\Phi_0^2r_0^2} - 2\sqrt{\ell^2 + \Phi_0^2r_0^2}\right],\\
	\Omega_d &= - 4N \sum_{\ell=1}^\infty \ell \sqrt{\ell^2 + \Phi_0^2r_0^2}.
\end{align}
Using Euler--Maclaurin, the regular part gives, at leading order,
\begin{equation}
	\Omega_r = -\frac{2 N}{3} (r_0\mu)^3 \pqty*{ 3(\kappa_0^2-1) \arccoth{\kappa_0} + 3\kappa_0 - 2 \kappa_0^3 + 2 (\kappa_0^2-1)^{\frac{3}{2}} }+\dots,
\end{equation}
while using zeta-function regularization, we find for the divergent part
\begin{equation}
	\Omega_d =  N \pqty*{ \frac{4(\Phi_0 r_0)^3}{3} + \frac{\Phi_0 r_0}{3} - \frac{1}{60 \Phi_0 r_0} + \dots } .
\end{equation}
Finally, we can obtain the relation $\mu = \mu (Q)$ by inverting order-by-order the charge equation in~\eqref{eq:NJL_scalingdim}. As already mentioned, the sum appearing there is convergent, so we can simply apply the Euler--Maclaurin formula. The result of the gap equation~\eqref{eq:ansatz} with the coefficients in~\eqref{eq:coeffs} together with the relation $\mu = \mu (Q)$ lead to the following asymptotic expansion for the scaling dimension $\Delta_{SF}$:
\begin{equation}
  \frac{\Delta_{SF}(Q)}{2N} = \frac{2}{3} \pqty*{\frac{Q}{2N \kappa_0}}^{3/2} + \frac{1}{6} \pqty*{\frac{Q}{2N \kappa_0}}^{1/2} + \frac{11-6\kappa_0^2}{720\kappa_0^{2}} \pqty*{\frac{Q}{2N \kappa_0}}^{-1/2} + \dots
\end{equation}
This is one of the main results of this work. Note that the leading term is consistent with the expression that we had found for the ground state energy in flat space, Eq.~\eqref{eq:flat_space_NJL}. We can think of the subleading terms in the large-charge expansion as of an expansion in the curvature, so that the leading term just depends on the volume of the manifold, which for the two-sphere is $V= 4\pi r_0^2$.

The conformal Goldstone identified in Sec.~\ref{sec:fluctuations} is present for any $N$ and gives rise to a universal contribution to the scaling dimension at order $N^0$, $Q^0$ with the same numerical value as for the \(O(N)\) model, see Eq.~\eqref{eq:scalingON}.
Moreover, its excitations generate the local spectrum on top of the primary $\mathcal{Q}_Q$.

\subsubsection{Small-Q regime}%
\label{sec:smallQ}

Next we want to study the limit of $Q/N \ll 1$.
At $Q=0$, the free energy has to vanish by conformal invariance as it corresponds to the conformal dimension of the identity operator.
In our description in terms of a composite scalar field \(\Phi\), we need to take into account its conformal coupling to the curved background.
For the three-dimensional cylinder this is the same as a mass term \(m = 1/(2r_0) \), which in our description comes from the chemical potential (see~\cite{Alvarez-Gaume:2019biu} for the analogous derivation in the large-N vector model).
A direct calculation shows that for $\mu=1/(2r_0)$ the free energy and the charge both vanish and the gap equation on the sphere is satisfied with a zero, \ac{vev} $\Phi_0=0$.
The small-charge expansion has to be done around this point, so it is convenient to write $\mu = 1/(2r_0) + \hat \mu$ and expand for small $\hat \mu$ all the expressions in Eq.~\eqref{eq:NJL_scalingdim}. At $\hat{\mu} = 0$ the symmetry is restored, so $\Phi_0$ does not acquire a \ac{vev}. We can then write an ansatz for $\Phi_0$ in the form
\begin{equation}
	\hat \mu = \mu_2 \Phi_0^2 r_0 + \mu_4 \Phi_0^4  r_0^3 + \dots .
 \label{eq:ansatz_2}
\end{equation}
The corresponding charge can be computed expanding in $\Phi_0 r_0 \ll 1$, and one finds
\begin{align}
	\frac{Q}{2N} =& \frac{\pi^2}{4} (\Phi_0r_0)^2 - \frac{\pi^2}{16}(\pi^2 - 16\mu_2) (\Phi_0r_0)^4 \nonumber\\
 &+ \frac{\pi^2}{48} \left( \pi^4 + 12 \pi^2 (\mu_2^2 - 2 \mu_2) + 48 \mu_4   \right) (\Phi_0 r_0)^6 + \dots,
\end{align}
which manifestly vanishes for \(\Phi_0 = 0\). The coefficients $\mu_{i}$ appearing in the ansatz \eqref{eq:ansatz_2} are determined solving the gap equation \eqref{eq:gap_NJL} order by order in small $\hat{\mu}$. Once again, we separate the gap equation into a divergent and a convergent sum as done before, both of them having a well-defined expansion for $\Phi_0 r_0 \ll 1$, and this time we subtract the $\hat{\mu} =0$ contribution. The convergent part is  
\begin{equation}
	G_r  = \frac{\pi^2}{2} \mu_2 (\Phi_0r_0)^2 - \frac{\pi^2}{4}(\pi^2\mu_2 - 4 \mu_2^2 - 2\mu_4) (\Phi_0r_0)^4+\dots,
\end{equation}
while the divergent part can be regularized order-by-order using $\zeta$-function regularization as follows:
\begin{align}
	G_d &= \sum_{\ell = 0}^\infty (2\ell+1)\frac{1}{\sqrt{(\ell +\tfrac{1}{2})^2+ (\Phi_0 r_0)^2}} \nonumber\\
    &= 2 \sum_{\ell=0} (\ell+\tfrac{1}{2}) \sum_{k=0} \binom{-\sfrac{1}{2}}{k}(\Phi_0r_0)^{2k}(\ell+\tfrac{1}{2})^{2(-1/2-k)} \nonumber \\
  &= 2 \sum_{k=0} \binom{-\sfrac{1}{2}}{k}(\Phi_0r_0)^{2k} \zeta(2k; \tfrac{1}{2}) \nonumber\\
	&= -\frac{\pi^2 (\Phi_0r_0)^2}{2} + \frac{\pi^4 (\Phi_0r_0)^4}{8} + \dots.	
\end{align}
Solving the gap equation $G_r + G_d = 0$ order by order, we can determine all the coefficients $\mu_i$. The first two are given by
\begin{align}
	\mu_2 &= 1, & \mu_4 &=\frac{\pi^2-8}{4}, &&\dots
\end{align}
As done in the previous section, we apply the same procedure to compute the divergent sum appearing in the scaling dimension~\eqref{eq:NJL_scalingdim}, which corresponds to the thermodynamic potential $\Omega/(2N)$. This quantity can be divided in the two contributions
\begin{align}
  &\begin{multlined}[][\arraycolsep]
    \Omega_r = -2N\sum_{\ell = 1} \ell 
    \left[ \sqrt{(\ell + \tfrac{1}{2} +\hat\mu r_0)^2+(\Phi_0 r_0)^2} + \sqrt{(\ell - \tfrac{1}{2} - \hat\mu r_0)^2+(\Phi_0 r_0)^2} \right.\\
    - \left.\sqrt{(\ell + \tfrac{1}{2})^2+(\Phi_0 r_0)^2} + \sqrt{(\ell - \tfrac{1}{2})^2+(\Phi_0 r_0)^2} \right],
  \end{multlined}
  \\
	&\Omega_d = - 4N \sum_{\ell=1}^\infty (\ell+ \tfrac{1}{2}) \sqrt{(\ell+\tfrac{1}{2})^2+ (\Phi_0 r_0)^2}.
\end{align}
The regular part can be expanded for $\Phi_0 r_0 \ll 1$ inside the sum, and at leading order one finds
\begin{equation}
\Omega_r = N \pqty*{ \frac{\pi^2}{2}\mu_2 (\Phi_0 r_0)^4 + \dots }.
\end{equation}
The divergent part can be regularized using $\zeta$-function regularization:
\begin{equation}
	\begin{aligned}
	\Omega_d &= -4 N \sum_{\ell = 0}^\infty (\ell+ \tfrac{1}{2}) \sum_{k=0} \binom{\sfrac{1}{2}}{k}(\Phi_0 r_0)^{2k} (\ell+\tfrac{1}{2})^{2(1/2-k)} \\
           & =- 4 N \sum_{k=0} \binom{\sfrac{1}{2}}{k}(\Phi_0 r_0)^{2k} \zeta(2k-2; \tfrac{1}{2}) \\
	&=  N \frac{\pi^2}{4}(r_0 \Phi_0)^4+\dots.
\end{aligned}
\end{equation}
Finally, we just need to invert $Q= Q(\Phi_0)$:
\begin{equation}
	r_0\Phi_0 = \frac{2}{\pi} \pqty*{\frac{Q}{2N}}^{1/2}+\frac{\pi^2-16}{\pi^3} \pqty*{\frac{Q}{2N} }^{3/2}+\dots
\end{equation}
to find that the conformal dimension in the limit of small charge is given by
\begin{equation}
	\frac{\Delta(Q)}{2N}  = \frac{1}{2}\left(\frac{Q}{2N}\right) +\frac{2}{\pi^2}\left(\frac{Q}{2N} \right)^2+ \dots
\end{equation}
As $\Phi$ has charge 2, the leading-order result $\Delta(Q) = Q/2$ is the expected relation for the operator $\Phi^{Q/2}$ in the free-field limit. Since to leading order, this relation is independent of $N$, this leading order result also applies to the $SU(2)_L \times SU(2)_R$ model.

\section{Conclusions}%
\label{sec:conclusions}

In this work we have studied several fermionic models with four-fermi interactions, namely the \ac{gn} model, the \ac{njl} model and its $SU(2)_L \times SU(2)_R$ generalization, in three dimensions in sectors of large charge and large N. 
Due to the fermionic nature of our models, we find that the large-charge ground state has two possible descriptions, depending on whether the model has a symmetry that lifts to an axial symmetry in four dimensions.
If none is present, as in the \ac{gn} model, the fixed-charge sector has a Fermi surface description. %
On the other hand, when fixing the axial charge, as can be done in the \ac{njl}-type models, we recover the superfluid description familiar from the bosonic case%
\footnote{In the $SU(2)$-\ac{njl} one can also fix the vectorial combination and still find a \ac{ssb} ground state}. 
For the $U(1)$-\ac{njl} model, the physics of the ground state becomes more transparent once we realize that it can be mapped to the Cooper model via a Pauli--Gürsey map: in the presence of an attractive interaction, the Cooper instability leads to the breakdown of the Fermi surface as Cooper pairs form and give rise to a bosonic condensate. 

For the cases with superfluid description, the predictions of the large-charge \ac{eft} are relevant. 
The condensate manifests itself in the large-\(N\) limit as a \ac{vev} for the collective field. 
Working in the large-\(N\) limit provided us with a controlled setting in which we could confirm the predictions of the superfluid \ac{eft}. 
Concretely, we have explicitly computed the fixed-charge ground state, the phonon spectrum over the ground-state and the scaling dimensions of the lowest operator at fixed charge.

\bigskip
Our large-\(N\) analysis of the large-charge sectors of several fermionic models constitutes a good starting point for a number of further directions:
\begin{itemize}
\item As we have already pointed out, the absence of \ac{ssb} in the large-\(N\) \ac{gn} model might either indicate that the physics is the one of a Fermi sphere at all energies, with a possible finite-\(N\) transition, or simply the fact that the \ac{ssb} is exponentially suppressed in \(1/N\) and hence invisible in perturbation theory. This is certainly something that needs to be understood, especially because the former alternative would be the first example of an interacting theory in which the large-charge physics is not described by a superfluid.
\item While we have worked mostly at leading order in \(N\), calculating the full $1/N$ corrections is the logical next step which we do no expect to present any conceptual obstacles for the models with leading-order \ac{ssb}.
\item A resurgence analysis of the superfluid cases along the lines of~\cite{Dondi:2021buw} would further complete our picture of the large-charge expansion of fermionic models.
    \item In this work we have studied the \ac{njl} model with only a fixed $U(1)_A$ or $U(1)_B$ charge. Studying the mixed phase with both finite $\mu_B$ and $\mu_A$ might be of interest, as it should interpolate between a Fermi sphere ground state and a superfluid one.
    \item We have studied explicitly the $SU(2)_L\times SU(2)_R$ generalization of the \ac{njl} model, but also its $SU(M)_L\times SU(M)_R$ generalization with $M \ll N$ could provide further insight in the study of non-abelian large charge sectors.
\end{itemize}

  \section*{Acknowledgments}

  \begin{small}
    \dosserif{\noindent{}We would like to thank Luis Álvarez--Gaumé, Oleg Antipin, John Gracey, Gerald Dunne and Giacomo Sberveglieri for illuminating discussions and Jahmall Bersini for a careful reading of the manuscript.
    The work of S.H. is supported by
    the World Premier International Research Center Initiative (WPI Initiative), MEXT, Japan;
    by the JSPS Program for Advancing Strategic International Networks to Accelerate the
    Circulation of Talented Researchers; and also supported in part by JSPJ KAKENHI Grant
    Numbers JP22740153, JP26400242.%
    The work of N.D., R.M. and S.R. is supported by the Swiss National Science Foundation under grant number 200021 192137.
    S.H. would like to thank the University of Bern and the SwissMAP research station for hospitality.
    S.H., D.O, and S.R. gratefully acknowledge support from the Simons Center for Geometry and Physics, Stony Brook University at which some of the research for this paper was performed.}
  \end{small}

\appendix
\section{3D Fermions}\label{sec:app_fermions}

In this appendix we collect some background material for fermionic theories in 2+1 and 3 dimensions in order to make the present work as self-contained as possible.
\subsection{Gamma matrices in the Dirac convention in 3D} \label{sec:notation}
The gamma matrices in 2+1 and 3 dimensions are built out of the Pauli matrices
 \begin{align}\label{Pauli}
 	\sigma_1 & = \begin{pmatrix}
 		0 & 1 \\ 1 & 0
 	\end{pmatrix}, & 
 	\sigma_2 &= \begin{pmatrix}
 		0 & -i \\ i & 0
 	\end{pmatrix}, &
 	\sigma_3 &=  \begin{pmatrix}
 		1 & 0 \\ 0 & -1
 	\end{pmatrix},
 \end{align}
 as follows:
 \begin{align}\label{eq:GammaMatricesConvention}
 d&=2+1:	& &\gamma_0 = i\sigma_3 , & &\gamma_{1,2} = \sigma_{1,2} \\
 d&=3:	 & &\gamma_\mu = \sigma_\mu, & &\mu = 1,2,3.
 \end{align}
And our convention for the Clifford algebra is
\begin{equation}
    \{ \gamma_\mu , \gamma_\nu \} = 2 \eta_{\mu \nu}.
\end{equation}
We have chosen the signature $\eta_{\mu\nu} = (-1,1,1)$ for $2+1$ dimensional Minkowski spacetime. In this signature $\gamma_0$ is anti-Hermitian, while spatial $\gamma$s are Hermitian.\\
Moreover, the gamma matrices satisfy
\begin{align}
(\gamma_i)^2 & = - (\gamma_0)^2 = \mathbbm{1}, & \gamma_0 \gamma_\mu \gamma_0 &= (\gamma_\mu)^\dagger.
\end{align}
Complex (Dirac) spinors $\psi$ transform in the usual representation of $SO(1,2),\, SO(3)$ generated by these gamma matrices.\\
The Dirac conjugate in our notation is

\begin{align}
   d& =2+1: & \bar{\psi}& = \psi^\dagger \gamma_0, \\
   d& =3: &\bar{\psi} &=\psi^\dagger \gamma_3.
\end{align}
The continuation from Minkowski to Euclidean spacetime is obtained as follows:

\begin{align}
    t  & \rightarrow -i \tau, &\partial_t &\rightarrow i \partial_\tau, &\gamma_0  &\rightarrow i \gamma_3, & \gamma_i &\rightarrow \gamma_i.
\end{align}
Given the above transformation rules, the massive Dirac action is continued in the following manner:

\begin{align}
 &i \underbrace{\int \dd t \dd^2 x \, \left [ \bar{\psi} ( i \gamma^\mu \del_\mu + i m )\psi\right] }_{S_M} &\longrightarrow& &-\underbrace{ \int \dd^3 x  \left[  \bar{\psi} (\gamma^\mu \del_\mu + m) \psi \right ] }_{S_E} .
 \end{align}

 \subsection{Spinors on $S^1_\beta \times S^2$}
\label{sec:cylinder_spinor}

\paragraph{Spinors in spherical coordinates.}
~\\
In this section we follow closely the treatment outlined \cite{Borokhov_2002}.\\
The Hermitian Dirac operator on $\mathbb{R}^3$ is
\begin{equation}
     i \gamma^\mu \del_\mu = - \vec{\sigma} \cdot \vec{p}
\end{equation}with momentum operator $p_\mu = - i \partial_\mu$ and $\vec{\sigma} = \sigma_i$ the Pauli matrices of Eq.~\eqref{Pauli}.\\
We define the generalized angular momentum and total angular momentum as

\begin{align}
	\vec{L} &= \vec{r} \times \vec{p} , & \vec{J} &= \vec{L} + \frac{\vec{\sigma}}{2} ,& [ \vec{L}, \vec{r} ] &= [\vec{J}, \vec{r}] = 0.
\end{align}
These are both Hermitian operators. The eigenfunctions of $\vec{L}^2$ are ordinary spherical harmonics:

\begin{align}
	\vec{L}^2 Y_{\ell m} &= \ell (\ell+1 ) Y_{\ell m} , & L_z Y_{\ell m} &= m Y_{\ell m}, & \ell &= 0,1,2... & m &= -\ell , ... \ell.
\end{align}
Using these, we can build simultaneous eigenfunctions of $\{ \vec{J}^2 ,  J_z, \vec{L}^2 , \vec{S}^2 \}$. These are spinor spherical harmonics:

\begin{align}
	\phi^+_{ j m_j} &= \begin{pmatrix}
	\sqrt{\frac{\ell+m+1}{2\ell+1}} Y_{\ell m} \\ \sqrt{\frac{\ell-m}{2\ell+1}} Y_{\ell m+1} 
\end{pmatrix}, &
\phi^-_{ j m_j} &= \begin{pmatrix}
- \sqrt{\frac{\ell-m}{2\ell+1}}  Y_{\ell m} \\ \sqrt{\frac{\ell+m+1}{2\ell+1}} Y_{\ell m+1} 
\end{pmatrix}.
\end{align}
The wave functions $\phi^{\pm}$ correspond respectively to the cases $j = \ell \pm 1/2$ and $m_j = m \pm 1/2$. These have the following quantum numbers:
\begin{align}
	&\begin{cases}
		\vec{L}^2 \phi^\pm_{j m_j} &= \ell(\ell+1) \phi^\pm_{j  m_j}\\
				\vec{J}^2 \phi^\pm_{j m_j} &= j(j+1) \phi^\pm_{j  m_j}\\
						J_z \phi^\pm_{j m_j} &= m_j \phi^\pm_{j  m_j}
	\end{cases} & \begin{cases}
	j =  \frac{1}{2} ,  \frac{3}{2} , \frac{5}{2} ... \\
	m_j = -j ... j
\end{cases}
\end{align}
and are $(2j+1)$-degenerate. Any spinor in $\mathbb{R}^3$ can be decomposed in this orthonormal basis. 
It is convenient to introduce the radial $\gamma$ matrix $\gamma_r = \vec{\gamma} \cdot \hat{r}$. The Dirac operator can then be written as
\begin{equation}
	i \gamma^\mu \del_\mu = i \sigma_r \left\{  \frac{\partial}{\partial r} - \frac{1}{r} \left[  \vec{J}^2 - \vec{L}^2 - \frac{3}{4}  \right]    \right\} ,
\end{equation}
and is diagonal in the $\phi^\pm$ basis.

\paragraph{Weyl map to the cylinder.}
~\\
We perform a Weyl transformation to the cylinder by letting

\begin{align}
    r&=e^{\tau}, & \eta_{\mu\nu}&=Re^{2 \tau} g_{\mu\nu}, & \psi_{\mathbb{R}^3}&=e^{-\tau} \psi_{\mathbb{R} \times S^2}.
\end{align}
If we foliate $\mathbb{R}^3$ radially we can define the Dirac conjugate as $\psi^\dagger = \bar{\psi} \sigma_r$. Then the free Dirac action on $\mathbb{R}^3$ reads
\begin{align}
	S &= \int_{\mathbb{R}^3}\, \bar{\psi}\slashed{\partial}\psi= \int_{\mathbb{R} \times S^2} \bar{\psi} \slashed{D} \psi, & \slashed{D} &= \gamma_r  \left\{  \frac{\partial}{\partial \tau} - \frac{1}{R} \left[ \hat{J}^2 - \hat{L}^2 + \frac{1}{4}  \right] \right\}.
\end{align}
The eigenfunctions on the cylinder are
\begin{equation}
	\Psi^\pm_{n j m_j}(\tau , \hat{x}) = e^{-i \omega_n \tau} \phi_{j m_j}^{\pm}(\hat{x}),
\end{equation}
where $\hat{x}$ is a point on $S^2$. We will make use of the following relations when computing functional determinant on $S_\beta^1 \times S^2$:
\begin{align}
	\int_{\mathbb{R} \times S^2} \, (\Psi^{\pm}_{j m_j})^\dagger  \Psi^{\pm}_{j' m_j'} &= \delta_{jj'} \delta_{m_j m_j'} \begin{pmatrix}
		1 & 0\\ 0 & 1
	\end{pmatrix}, \\
	\int_{\mathbb{R} \times S^2} \, (\Psi^{\pm}_{j m_j})^\dagger \gamma_r \Psi^{\pm}_{j' m_j'} &= \delta_{jj'} \delta_{m_j m_j'} \begin{pmatrix}
		0 & -1\\ - 1 & 0
	\end{pmatrix}, \\
  	\int_{\mathbb{R} \times S^2} \, (\Psi^{\pm}_{j m_j})^\dagger (i \slashed{D}) \Psi^{\pm}_{j' m_j'} &= \delta_{jj'} \delta_{m_j m_j'} \begin{pmatrix}
		0 & \omega_n -i \omega_j\\  \omega_n+ i \omega_j & 0
	\end{pmatrix},
\end{align}
where we introduced
\begin{align}
	\omega_n &= \frac{(2n+1)\pi}{\beta}, & \omega_j &= \frac{1}{R} \left( j + \frac{1}{2} \right),
\end{align}
which are the Matsubara frequencies and the eigenvalues of the Dirac operator on the sphere, respectively.

\subsection{Reducible Representation} 
\label{sec:RedRep}

For three-dimensional fermionic theories with an even number $2N$ of fermion fields $\psi_{a=1...2N}$ it is convenient to introduce a reducible representation of the Clifford algebra as follows:
 \begin{align}
 	\Gamma_\mu &= \sigma_3 \otimes \gamma_\mu = \begin{pmatrix}
 	\gamma_\mu & 0 \\ 0 & - \gamma_\mu 
 	\end{pmatrix}, &   \Psi_a &\equiv  \begin{pmatrix}
 	\psi_a \\ \psi_{a+N}
 \end{pmatrix}, & a = 1, ... , N.
 \end{align}
then we can pick
\begin{equation}
  \Gamma_5 = \sigma_1 \otimes \mathbbm{1} =
  \begin{pmatrix}
    & \mathbbm{1} \\ \mathbbm{1} 
  \end{pmatrix}.
\end{equation}
The charge conjugation matrix is 
\begin{equation}
    C_4 = \Gamma_2 = \sigma_3 \otimes C = \begin{pmatrix}
     \sigma_2 &  \\
       & -\sigma_2
    \end{pmatrix} ,
\end{equation}
and satisfies
\begin{align}
    &C_4 = C_4^{-1} = C_4^\dagger = - C_4^T = - C_4^*, \quad &C_4 \Gamma_\mu C_4 = - (\Gamma_\mu)^T .
\end{align}
Charge conjugation is independent on the signature of space~\cite{Wetterich:2011ab}.
In addition, it holds that
\begin{align}
    \{ \Gamma_5 ,C_4\} =0 .
\end{align}
In terms of spinors the reducible four-dimensional representation consists of two two-dimensional irreducible spinors,
\begin{align}
    &\Psi = \big( \psi_{1} , \psi_{2} \big)^T , \quad &\bar \Psi = \Psi^\dagger \Gamma_3 =  \big( \psi_{1}^\dagger \gamma_3 , - \psi_{2}^\dagger \gamma_3 \big) = \big( \bar{\psi}_{1} , - \bar{\psi}_{2} \big) ,
\end{align}
for $a=1,\dots,N$. \\
As a concrete example, the action of the $U(1)$ \ac{njl} model in terms of this reducible representation can be written as
\begin{equation}
  S =  \int \dd^3x \left( \bar \Psi \Gamma^\mu \del_\mu \Psi - \frac{ g}{N} \left( (\bar \Psi \Psi)^2 - ( \bar \Psi \Gamma_5 \Psi)^2 \right) \right) .
\end{equation}

\section{U(1) Pauli--Gürsey transformation}
\label{sec:Pauli-Gursey}

We consider the following transformation at the level of the path integral:
\begin{equation}
    \begin{aligned}
        &\Psi \mapsto \frac{ 1}{2} \left[ (1-\Gamma_5) \Psi + (1+ \Gamma_5) C_4 \bar\Psi^T  \right] 
        , \\
        &\bar\Psi \mapsto 
        \frac{ 1}{2} \left[ \bar\Psi (1+\Gamma_5) - \Psi^T C_4 (1- \Gamma_5) \right] %
        .
    \end{aligned}
\end{equation}
Note that the precise form of the transformation depends on the convention for the gamma matrices.
This is an involution, it maps $\Psi_a$ to $\Psi_a$ after twice applying the transformation.
Under the \ac{pg} transformation the kinetic term remains invariant,
\begin{align}
   & \int \dd^3 x \, \bar \Psi \Gamma^\mu \partial_\mu \Psi \,\, \mapsto \,\, \int \dd^3 x \, \bar \Psi  \Gamma^\mu \partial_\mu \Psi ,
\end{align}
and the Cooper \ac{bcs} interaction term is mapped to the chiral \ac{gn} interaction term,
\begin{equation}
    \begin{aligned}
        &&- \bar \Psi C_4 \bar \Psi^T \, \Psi^T C_4 \Psi \mapsto \bar \Psi (1+ \Gamma_5) \Psi \, \bar \Psi (1- \Gamma_5) \Psi .
    \end{aligned}
\end{equation}
The converse statement is of course true as well, which is shown either by the fact that the \ac{pg} transformation is an involution, or directly by computing it. In the latter approach one will need the fact that (in our convention)
\begin{equation}
    \bar \Psi \Gamma_5 C_4 \bar \Psi^T = \Psi^T \Gamma_5 C_4 \Psi = 0 .
\end{equation}
Finally, the \ac{pg} transformation maps the fermion number chemical potential into the chiral (axial) chemical potential and vice versa,
\begin{align}
        \bar \Psi \Gamma_3 \mu \Psi &\mapsto \bar \Psi ( - \Gamma_3 \Gamma_5 \mu ) \Psi , & \bar \Psi \Gamma_3 \Gamma_5 \mu \Psi &\mapsto \bar \Psi ( - \Gamma_3 \mu ) \Psi .
    \end{align}
In total, we have the following map for the $U(1)$ \ac{njl} model:
\begin{equation}
    \begin{aligned}
        S &= \int \dd^3x \left( \bar \Psi ( \Gamma^\mu \del_\mu - \mu \Gamma_3 \Gamma_5 ) \Psi - \frac{ g}{N} \left( (\bar \Psi \Psi)^2 - ( \bar \Psi \Gamma_5 \Psi)^2  \right) \right) \\
        &= \int \dd^3x \left( \bar \Psi ( \Gamma^\mu \del_\mu - \mu \Gamma_3 \Gamma_5 ) \Psi - \frac{ g}{N} \bar \Psi ( 1 + \Gamma_5 ) \Psi \, \bar \Psi ( 1 - \Gamma_5 ) \Psi \right) \\
        &\mapsto \int \dd^3x \left( \bar \Psi ( \Gamma^\mu \del_\mu + \mu \Gamma_3 ) \Psi + \frac{ g}{N} \bar \Psi C_4 \bar \Psi^T \, \Psi^T C_4 \Psi \right) .
    \end{aligned}
\end{equation}
For completeness, written in terms of collective fields the Cooper model reads
\begin{equation}
  L =   \bar \Psi \Gamma^\mu \del_\mu \Psi + i \frac{\Phi}{2} \bar \Psi C_4 \bar \Psi^T + i \frac{ \Phi^*}{ 2} \Psi^T C_4 \Psi + \frac{ N}{4g} \Phi^{*} \Phi .
\end{equation}

\section{Finite-density loop integrals and Matsubara sums}
\label{sec:loop}

\subsubsection{Fourier transforms on $S^1_\beta \times \mathbb{R}^2$.}
	
We denote a point on $S^1_\beta \times \mathbb{R}^2$ as $X = (\tau, x)$ and momenta as $P = (\omega_n , \vec{p})$ where $\omega_n = \pi (2n+1)/\beta$ are fermionic Matsubara frequencies. Our normalization conventions for Fourier transforms are
\begin{align}
\delta(X-X') &=\SumInt \frac{\dd^d p}{\beta (2\pi)^d} e^{- i P \cdot (X-X')} & \delta_{nn'} \delta(p-p') &= \int \frac{\dd \tau \dd^d x}{\beta (2\pi)^d} e^{- i X \cdot (P-P')} \\
f(X) &=\SumInt \frac{\dd^d p}{\sqrt{\beta(2\pi)^d}} e^{- i P \cdot X} f(P), & f(P) &= \int \frac{\dd \tau \dd^d x}{\sqrt{\beta (2\pi)^d}}  e^{i P \cdot X} f(X).
\end{align}
\subsubsection{Matsubara sums}

The master formula for fermionic Matsubara sums encountered in Dirac determinant computations is
\begin{equation}
\sum_{n \in \mathbb{Z}} \log \left[ \frac{(2n+1)^2 \pi^2 + A^2}{(2n+1)^2 \pi^2 + 1 } \right] = A + 2 \log \left( 1 + e^{- A} \right).
\end{equation}
\subsection{GN scalar integrals at finite $\mu,\beta$ }

In this appendix we collect the scalar integral used to obtain the results in Section~\ref{sec:gn-fluctuations}.
Following~\cite{Laine:2016hma}, one-loop integrals at finite temperature and chemical potential can all be derived from the massive scalar integral
\begin{equation}
	\int \frac{\dd^d k}{(2\pi)^d} \frac{1}{[k^2+m^2]^\alpha} = \frac{1}{(4\pi)^{\frac{d}{2}}} \frac{\Gamma(\alpha-d/2)}{\Gamma(\alpha)} (m^2)^{-\alpha + \frac{d}{2}} .
\end{equation}
Recalling that $\tilde{K} = ( \omega_n - I \mu , \vec{k})$ the first scalar integral we used can be computed as
\begin{align}
	I_1 &=  \SumInt \frac{\dd^d k}{\beta (2\pi)^d} \frac{1}{\tilde{K}^2} \nonumber\\
	&= \frac{\Gamma(1-d/2)}{(4\pi)^\frac{d}{2}} \sum_{n \in \mathbb{Z}} \frac{1}{[(\omega_n - i \mu)^2]^{1 - \frac{d}{2}}} \nonumber\\
	&=\frac{\Gamma(1-d/2)}{(4\pi)^\frac{d}{2}} \left( \frac{2\pi}{\beta} \right)^{-2 + d} \sum_{n \in \mathbb{Z}} \frac{1}{\left[ \left( n + \frac{1}{2} - i \bar{\mu} \right)^2 \right]^{1 - \frac{d}{2}} }\nonumber\\
	&=\frac{\Gamma(1-d/2)}{(4\pi)^\frac{d}{2}} \left( \frac{2\pi}{\beta} \right)^{-2 + d} \left\{    \zeta\left( 2 - d , \frac{1}{2} - i \bar{\mu} \right) +   \zeta\left( 2 - d , \frac{1}{2} + i \bar{\mu} \right)   \right\} .
\end{align}
Barred quantities are normalized as $\bar{\mu} = \beta \mu/(2\pi)$ etc. At zero temperature this becomes
\begin{equation}
	\lim_{\beta \rightarrow \infty}  I_1 = - \frac{\mu}{4\pi} .
\end{equation}
The $I_2$ integral has three scales: $\beta,\mu, P$ where $P$ is an external momentum. It can be computed similarly to $I_1$ where a Feynman parametrization is used to merge the two propagators:
\begin{align}
	I_2 &=\SumInt \frac{\dd^2 k \dd^2 q}{\beta (2\pi)^2} \frac{\delta(K+Q-P)}{\tilde{K}^2 (\tilde{Q}^\dagger)^2 } \nonumber\\
	&= \int_0^1 \dd x \SumInt \frac{\dd^2 k }{\beta (2\pi)^2} \frac{1}{ \left[ k^2 + \left\{   x(1-x) p^2 + (1-x)(\omega_n - i \mu)^2 + x (\omega_m - \omega_n + i \mu)^2   \right\} \right]^2 }\nonumber \\
	&= \frac{\Gamma(2-d/2)}{\beta (4\pi)^{\frac{d}{2}}} \left( \frac{2\pi}{\beta} \right)^{d-4} \int_0^1 \dd x \sum_{n \in \mathbb{Z}}  \frac{1}{ \left[ (n + \frac{1}{2}  - i \bar{\mu} - x \bar{\omega}_m )^2 + x(1-x) (\bar{p}^2 + \bar{\omega}^2)   \right]^{2- \frac{d}{2}} }\nonumber\\
	&= \frac{\Gamma(2-d/2)}{\beta (4\pi)^{\frac{d}{2}}} \left( \frac{2\pi}{\beta} \right)^{d-4} \int_0^1 \dd x \sum_{n \in \mathbb{Z}}  \frac{1}{ \left[ (n +A)^2 +B  \right]^{2- \frac{d}{2}} }\nonumber\\
	&= \frac{\Gamma(2-d/2)}{\beta (4\pi)^{\frac{d}{2}}} \left( \frac{2\pi}{\beta} \right)^{d-4} \int_0^1 \dd x \left\{ \frac{1}{[A^2 + B]^{2-\frac{d}{2}}} +F(2-d/2 ; A,B) +F(2-d/2 ; -A,B) \right\},
\end{align}
where we introduced 
\begin{equation}
	A = \frac{1}{2} - i \mu  - x \omega_m, \quad B = x(1-x) [\bar{p}^2 +\bar{\omega}_m^2 ].
\end{equation}
The functions $F$ are special $\zeta$-functions found in \cite{Elizalde:1995hck}. At zero temperature we find
\begin{equation}
	\lim_{\beta \rightarrow \infty} I_2 = \frac{1}{8\sqrt{\omega_m^2 + p^2}} = \frac{1}{8 \sqrt{P^2}} .
\end{equation}
\subsection{NJL loop integrals} \label{sec:njl-computations}

We want to compute the following integrals:
\begin{align}
	D^{-1}_{\sigma\sigma}(P) &= -\int \frac{\dd^3k}{(2\pi)^3} \Tr\left[D^{(\mu,\sigma)}(K)D^{(-\mu,-\sigma)}(P-K) \right],\\
	D^{-1}_{\sigma\pi}(P) &= -i\int \frac{\dd^3k}{(2\pi)^3}\Tr\left[D^{(\mu,\sigma)}(K)\Gamma_5 D^{(-\mu,-\sigma)}(P-K) \right],\\
	D^{-1}_{\pi\sigma}(P) &= -i\int \frac{\dd^3k}{(2\pi)^3}\Tr\left[ \Gamma_5 D^{(\mu,\sigma)}(K) D^{(-\mu,-\sigma)}(P-K) \right] ,\\
	D^{-1}_{\pi\pi}(P) &= \int \frac{\dd^3k}{(2\pi)^3}\Tr\left[\Gamma_5 D^{(\mu,\sigma)}(K)\Gamma_5 D^{(-\mu,-\sigma)}(P-K) \right],
\end{align}
where the fermion propagator is given by
\begin{equation}
\begin{split}
	D^{(\mu,\sigma)}( P) & = (-i\slashed{P} +\Phi_0 -\mu\Gamma_3\Gamma_5)^{-1}\\
	& = \frac{
    \left( \omega^2 + k^2 + \Phi_0^2 - \mu^2 + 2 \mu (i \omega \Gamma_3 + \Phi_0 ) \Gamma_3 \Gamma_5 \right) 
    }{\left(\omega^2 + \Phi_0^2 + (\mu + k)^2 \right)\left(\omega^2 + \Phi_0^2 + (\mu - k)^2 \right)} \left( i\slashed{P} +\Phi_0 -\mu\Gamma_3 \Gamma_5\right) .
\end{split}
\end{equation}

\subsubsection{Zeroth order in $P /\mu$}
At zeroth order in $P/\mu$ these integrals are now
\begin{align}
	 D^{-1}_{\sigma\sigma}(P)|_{\mathcal{O}(0)} & =\int \frac{\dd^2 k \dd k_0}{(2 \pi)^3}   \Bigg[ 4 \Phi_0^2 \left(\frac{1}{\left((k+\mu )^2+k_0^2+\Phi_0 ^2\right)^2}+\frac{1}{\left((k-\mu )^2+k_0^2+\Phi_0 ^2\right)^2}\right)  \nonumber \\
	 &-\frac{2}{(k-\mu )^2+k_0^2+\Phi_0 ^2}-\frac{2}{(k+\mu )^2+k_0^2+\Phi ^2} \Bigg], \\
	D^{-1}_{\sigma\pi}(P)|_{\mathcal{O}(0)} & =0,\\
	D^{-1}_{\pi\sigma}(P)|_{\mathcal{O}(0)} & =0,\\
	D^{-1}_{\pi\pi}(P)|_{\mathcal{O}(0)} & =- \int \frac{\dd^2 k \dd k_0}{(2 \pi)^3} \left[\frac{2}{(k+\mu )^2+k_0^2+\Phi_0 ^2}+\frac{2}{(k-\mu )^2+k_0^2+\Phi_0 ^2}\right].
\end{align}
We then perform the residue integral over $k_0$,
\begin{align}
     D^{-1}_{\sigma\sigma}(P)|_{\mathcal{O}(0)} & =- \int \frac{\dd^2 k }{(2 \pi)^3} \left[\frac{2 \pi  (k-\mu )^2}{\left((k-\mu )^2+\Phi_0 ^2\right)^{3/2}}+\frac{2 \pi  (k+\mu )^2}{\left((k+\mu )^2+\Phi_0 ^2\right)^{3/2}} \right], \\
	D^{-1}_{\pi\pi}(P)|_{\mathcal{O}(0)} & = \int \frac{\dd^2 k }{(2 \pi)^3} \left[2 \pi  \left(\frac{1}{\sqrt{(k+\mu )^2+\Phi_0 ^2}}+\frac{1}{\sqrt{(k-\mu )^2+\Phi_0 ^2}}\right) \right].
\end{align}
The remaining integrals are divergent. The divergence is however independent of $\mu$, so we can simply subtract the expression for $\mu=0$ to regularize them:
\begin{multline}
  D^{-1}_{\sigma\sigma}(P)|_{\mathcal{O}(0)} = \int \frac{\dd^2 k }{(2 \pi)^3} \left[ \frac{4 \pi  k^2}{\left(k^2+\Phi_0 ^2\right){}^{3/2}}-\frac{2 \pi  (k-\mu )^2}{\left((k-\mu )^2+\Phi _0^2\right){}^{3/2}}-\frac{2 \pi  (k+\mu )^2}{\left((k+\mu )^2+\Phi _0^2\right){}^{3/2}} \right] 
                           \\ - \int \frac{\dd^2 k }{(2 \pi)^3} \frac{4 \pi  k^2}{\left(k^2+\Phi_0 ^2\right){}^{3/2}}
\end{multline}
\begin{multline}                           
	D^{-1}_{\pi\pi}(P)|_{\mathcal{O}(0)}  = \int \frac{\dd^2 k }{(2 \pi)^3} \left[-\frac{ 4 \pi}{ \sqrt{ k^2+ \Phi_0 ^2}} +\frac{2 \pi}{ \sqrt{ (k-\mu )^2+\Phi_0 ^2}} +\frac{2 \pi}{ \sqrt{ (k+\mu )^2+\Phi_0^2}} \right] \\+ 4 \pi \int \frac{\dd^2 k }{(2 \pi)^3} \frac{1}{\sqrt{k^2+\Phi_0^2}}.
\end{multline}
The divergent integrals after regulation give
\begin{align}
    &\int \frac{\dd^2 k }{(2 \pi)^3} \frac{4 \pi  k^2}{\left(k^2+\Phi_0 ^2\right){}^{3/2}}  = \frac{8}{(2 \pi)^3} \pi^2 \int \dd k \ k \frac{k^2}{\left(k^2+\Phi _0^2\right){}^{3/2}} \equiv - \frac{16}{(2 \pi)^3} \pi^2 \Phi_0,\\
     &\int \frac{\dd^2 k }{(2 \pi)^3} \frac{4 \pi}{\sqrt{k^2+\Phi_0^2}} \equiv - \frac{8}{(2 \pi)^3} \pi^2 \Phi_0.
\end{align}
We can then perform the spatial integral over the momentum $k$. We get
\begin{align}
    D^{-1}_{\sigma\sigma}(P)|_{\mathcal{O}(0)} & = \frac{8 \pi ^2}{(2 \pi)^3} \left(2 \sqrt{\mu ^2+\Phi_0 ^2}-\mu  \arctanh\left(\tfrac{\mu }{\sqrt{\mu ^2+\Phi_0 ^2}}\right)\right), \\
    D^{-1}_{\pi\pi}(P)|_{\mathcal{O}(0)} & = \frac{8 \pi ^2}{(2 \pi)^3} \left(\sqrt{\mu ^2+\Phi_0 ^2}-\mu  \arctanh\left(\tfrac{\mu }{\sqrt{\mu ^2+\Phi_0 ^2}}\right)\right).
\end{align}
Finally we can use the \ac{eom},
\begin{align}
	\Phi_0 &=\sqrt{\kappa_0^2 -1}\mu, &
	\arctanh*(\tfrac{1}{\kappa_0}) &=\kappa_0
\end{align}
to find the final result
\begin{align}
    D^{-1}_{\sigma\sigma}(P)|_{\mathcal{O}(0)} & = \frac{\kappa_0 \pi}{\mu}, & 
    D^{-1}_{\pi\pi}(P)|_{\mathcal{O}(0)} & =0.
\end{align}

\subsubsection{First order in $P/ \mu$}

At order 1 in $P/\mu$ the following two integrals are an odd function of $k_0$ and $k_1,k_2$, hence under integration it follows that
\begin{align}
D^{-1}_{\sigma\sigma}(P)|_{\mathcal{O}(P/ \mu)} & = 0,\\ D^{-1}_{\pi\pi}(P)|_{\order{P/\mu}} & =0.
\end{align}
The remaining two are computed as follows:
\begin{align}
    D^{-1}_{\sigma\pi}(P) |_{\order{P/\mu}}& = \int \frac{\dd^2 k \dd k_0}{(2\pi)^3} \frac{4 \mu ^2 \frac{\omega}{\mu} \left(-3 k^4+2 k^2 \left(-k_0^2+\mu ^2-\Phi_0 ^2\right)+\left(k_0^2+\mu ^2+\Phi_0 ^2\right)^2\right)}{\left((k-\mu )^2+k_0^2+\Phi_0 ^2\right)^2 \left((k+\mu )^2+k_0^2+\Phi_0 ^2\right)^2}, \\
    D^{-1}_{\pi\sigma}(P) |_{\order{P/\mu}} & = -\int \frac{\dd^2 k \dd k_0}{(2\pi)^3}\frac{4 \mu ^2 \frac{\omega}{\mu}    \left(-3 k^4+2 k^2 \left(-k_0^2+\mu ^2-\Phi_0 ^2\right)+\left(k_0^2+\mu ^2+\Phi_0 ^2\right)^2\right)}{\left((k-\mu )^2+k_0^2+\Phi_0 ^2\right)^2 \left((k+\mu )^2+k_0^2+\Phi_0 ^2\right)^2}.
\end{align}
We perform the integral over $k_0$ to obtain
\begin{align}
    D^{-1}_{\sigma\pi}(P) |_{\order{P/\mu}}& = \pi \int \frac{\dd^2 k}{(2 \pi)^3}   \left(\frac{\mu  (\mu -k)}{\left((k-\mu )^2+\Phi_0 ^2\right)^{3/2}}+\frac{\mu  (k+\mu )}{\left((k+\mu )^2+\Phi_0 ^2\right)^{3/2}}\right) \frac{\omega}{\mu} ,\\
    D^{-1}_{\pi\sigma}(P) |_{\order{P/\mu}} & = \pi \int \frac{\dd^2 k}{(2 \pi)^3}    \left(\frac{\mu  (k-\mu )}{\left((k-\mu )^2+\Phi_0 ^2\right)^{3/2}}-\frac{\mu  (k+\mu )}{\left((k+\mu )^2+\Phi_0 ^2\right)^{3/2}}\right)\frac{\omega}{\mu} .
\end{align}
We can then perform the spatial integral over the momentum $k_1$ and $k_2$ 
\begin{align}
     D^{-1}_{\sigma\pi}(P) |_{\order{P/\mu}}& = \frac{2 \pi ^2}{(2 \pi)^3} \mu \frac{\omega}{\mu}   \log \left(\tfrac{2 \mu  \left(\mu -\sqrt{\mu ^2+\Phi_0 ^2}\right)}{\Phi_0 ^2}+1\right) ,\\
     D^{-1}_{\pi\sigma}(P) |_{\order{P/\mu}} & = \frac{2 \pi ^2}{(2 \pi)^3} \mu \frac{\omega}{\mu}  \log \left(\tfrac{2 \mu  \left(\sqrt{\mu ^2+\Phi_0 ^2}+\mu \right)}{\Phi_0 ^2}+1\right) .
\end{align}
Finally using the \ac{eom} and simplifying we end up with
\begin{align}
    D^{-1}_{\sigma\pi}(P) |_{\order{P/\mu}}& =  -\frac{\kappa_0  \omega }{2 \pi }, &  D^{-1}_{\pi\sigma}(P) |_{\order{P/\mu}} & = \frac{\kappa_0  \omega }{2 \pi }.
\end{align}

\subsubsection{Second order in $P^2 / \mu^2$}

Next we consider the quadratic part in $P$. By rotational invariance, at second order in $\mathcal{O}(P^2/\mu^2)$, the integrand must have the form
\begin{equation}
	A(k)\omega^2+ B(k)\omega(P\cdot k) + C(k)P^2+ D(k) (k\cdot P)^2.
\end{equation}
The $B(k)$ piece does not contribute as it is an odd function of $k_1,k_2$. Similarly, the cross-term in $(k\cdot P)^2$ will not contribute. The part of the integrand that contributes is thus
\begin{equation}
	A(k)\omega^2+  C(k)P^2+ D(k) (k_1^2p_1^2+k_2^2p_2^2).
\end{equation} 
Given the symmetry under the exchange $1 \leftrightarrow 2$, this is a function of $p_1^2+p_2^2$.\\
We will split the computation into two parts, one with $\omega$ and the other with $P$.\\
We start with the $\omega$ part. After performing the $k_0$ residue integral the two remaining integrals are
\begin{align}
    I^{\omega}_{\sigma \sigma} & = \frac{1}{2} \int \frac{\dd^2 k}{(2 \pi)^3} \pi  \mu ^2 \left (\frac{\omega}{\mu} \right)^2 \left(\frac{(k-\mu )^2}{\left((k-\mu )^2+\Phi_0 ^2\right)^{5/2}}+\frac{(k+\mu )^2}{\left((k+\mu )^2+\Phi_0 ^2\right)^{5/2}}\right) , \\
    I^{\omega}_{\pi \pi} & = \int \frac{\dd^2 k}{(2 \pi)^3} \frac{1}{2} \pi  \mu ^2 \left (\frac{\omega}{\mu} \right)^2 \left(\frac{1}{\left((k+\mu )^2+\Phi_0 ^2\right)^{3/2}}+\frac{1}{\left((k-\mu )^2+\Phi_0 ^2\right)^{3/2}}\right).
\end{align}
We can then perform the integrals over $k_1$ and $k_2$ to get
\begin{align}
     I^{\omega}_{\sigma \sigma} & = \frac{1}{(2 \pi)^3} \left (\frac{\omega}{\mu} \right)^2 \frac{2 \pi ^2 \mu ^2  \left(\mu ^2+2 \Phi_0 ^2\right)}{3 \Phi_0 ^2 \sqrt{\mu ^2+\Phi_0 ^2}}, &
     I^{\omega}_{\pi \pi}  & = \frac{1}{(2 \pi)^3} \left (\frac{\omega}{\mu} \right)^2\frac{2 \pi ^2 \mu ^2  \sqrt{\mu ^2+\Phi_0 ^2}}{\Phi_0 ^2}.
\end{align}
By using the \ac{eom} and simplifying we finally get
\begin{align}
   I^{\omega}_{\sigma \sigma}& = \frac{\omega ^2-2 \kappa_0 ^2 \omega ^2}{12 \pi  \kappa_0  \mu -12 \pi  \kappa_0 ^3 \mu }, &
   I^{\omega}_{\pi \pi}  & = -\frac{\kappa_0  \omega ^2}{4 \pi  \mu -4 \pi  \kappa_0 ^2 \mu }.
\end{align}
Similarly we can repeat the same procedure for the parts that depend on $P, \ p_1, \ p_2$. First we perform the residue integral over $k_0$ and then the integrals over $k_1, \ k_2$. The result is
\begin{align}
    I^P_{\sigma \sigma} & = \frac{1}{2} \frac{\pi ^2 \mu}{(2 \pi)^3} \left( \frac{p}{\mu} \right)^2 \left(2 \arctanh\left(\tfrac{\mu }{\sqrt{\mu ^2+\Phi_0 ^2}}\right)+\frac{2 \mu  \left(\mu ^4+\Phi_0 ^4\right)}{3 \Phi_0 ^2 \left(\mu ^2+\Phi_0 ^2\right)^{3/2}}\right),\\
    I^P_{\pi \pi} & = \frac{\pi ^2 \mu}{(2 \pi)^3}  \left( \frac{p}{\mu} \right)^2 \left(\frac{\mu  \sqrt{\mu ^2+\sigma ^2}}{\Phi_0 ^2}+\arctanh\left(\tfrac{\mu }{\sqrt{\mu ^2+\Phi_0 ^2}}\right)\right).
\end{align}
Again, by using the \ac{eom} and simplifying we finally get
\begin{align}
   I^P_{\sigma \sigma} & =\frac{\left(3 \kappa_0 ^6-2 \kappa_0 ^4-2 \kappa_0 ^2+2\right) p^2}{24 \pi  \kappa_0 ^3 \left(\kappa_0 ^2-1\right) \mu }, &
   I^P_{\pi \pi}  & =-\frac{\kappa_0 ^3 p^2}{8 \pi  \mu -8 \pi  \kappa_0 ^2 \mu }.
\end{align}
So finally we can put it all together to get
\begin{align}
    D^{-1}_{\sigma\sigma}(P)|_{\mathcal{O}(P^2/ \mu^2)} & = \frac{\omega ^2-2 \kappa_0 ^2 \omega ^2}{12 \pi  \kappa_0  \mu -12 \pi  \kappa_0 ^3 \mu } + \frac{\left(3 \kappa_0 ^6-2 \kappa_0 ^4-2 \kappa_0 ^2+2\right) p^2}{24 \pi  \kappa_0 ^3 \left(\kappa_0 ^2-1\right) \mu },\\
    D^{-1}_{\pi\pi}(P)|_{\order{P^2/\mu^2}} & =   -\frac{\kappa_0  \omega ^2}{4 \pi  \mu -4 \pi  \kappa_0 ^2 \mu } -\frac{\kappa_0 ^3 p^2}{8 \pi  \mu -8 \pi  \kappa_0 ^2 \mu }.
\end{align}

\setstretch{1}

\printbibliography{}

\end{document}